\documentclass[twocolumn,english,prb,amsmath,amssymb,eqsecnum,preprintnumbers]{revtex4}
\usepackage[T1]{fontenc}
\usepackage[latin9]{inputenc}
\usepackage{amstext}
\usepackage{amsmath}
\usepackage{amsfonts}
\usepackage{amssymb}
\usepackage{mathtools}
\usepackage{graphicx}
\usepackage{rotating}
\usepackage {bm}
\usepackage[section]{placeins}
\usepackage{color}
\usepackage{multirow}
\usepackage{cellspace}
\makeatletter
\@ifundefined{textcolor}{}
{%
 \definecolor{BLACK}{gray}{0}
 \definecolor{WHITE}{gray}{1}
 \definecolor{RED}{rgb}{1,0,0}
 \definecolor{GREEN}{rgb}{0,1,0}
 \definecolor{BLUE}{rgb}{0,0,1}
 \definecolor{CYAN}{cmyk}{1,0,0,0}
 \definecolor{MAGENTA}{cmyk}{0,1,0,0}
 \definecolor{YELLOW}{cmyk}{0,0,1,0}
}

\def\kF{k_{\text{F}}}
\def\vF{v_{\text{F}}}
\def\NF{N_{\text{F}}}

\def\chis{\chi_{\text{s}}}

\def\epsilonF{\epsilon_{\text{F}}}

\def\sgn{{\text{sgn\,}}}

\def\be{\begin{equation}}
\def\ee{\end{equation}}
\def\bea{\begin{eqnarray}}
\def\eea{\end{eqnarray}}
\def\bse{\begin{subequations}}
\def\ese{\end{subequations}}

\usepackage{dcolumn}
\usepackage{bm}
\usepackage{amsfonts}
\draft

\makeatother

\usepackage{babel}
\begin{document}
\preprint{arXiv:xxxx.xxxx}
\bigskip
%

\title{\mbox{} \\ Magnetic Quantum Phase Transitions in a Clean Dirac Metal}

\author{ D. Belitz$^{1,2}$ and T. R. Kirkpatrick$^{3}$}

\affiliation{$^{1}$ Department of Physics and Institute of Theoretical Science, University of Oregon, Eugene, OR 97403, USA\\
                 $^{2}$ Materials Science Institute, University of Oregon, Eugene, OR 97403, USA\\
                 $^{3}$ Institute for Physical Science and Technology, University of Maryland, College Park, MD 20742, USA }

\date{\today}
\begin{abstract}
We consider clean Dirac metals where the linear band crossing is caused by a strong spin-orbit interaction, and study the
quantum phase transitions from the paramagnetic phase to various magnetic phases,
including homogeneous ferromagnets, ferrimagnets, canted ferromagnets, and magnetic nematics. We show that in all of these
cases the coupling of fermionic soft modes to the order parameter generically renders the quantum phase transition first order,
with certain gapless Dirac systems providing a possible exception. These results are surprising since a strong spin-orbit scattering 
suppresses the mechanism that causes the first order transition in ordinary metals. The important role of chirality in generating a 
new mechanism for a first-order transition is stressed.
\end{abstract}
%
%
\maketitle

\section{Introduction}
\label{sec:I}

It has been known for a long time that the spin-orbit interaction can lead to semimetals, that is, materials that in a well-defined sense are in between
metals and insulators.\cite{Herring_1937, Abrikosov_Beneslavskii_1970, Landau_Lifshitz_IX_1991} A Dirac semimetal is realized if two doubly degenerate
bands cross in one point in momentum space and the Fermi energy is at the crossing point, i.e., the valence band is full, the conduction
band is empty, and the Fermi `surface' consists of a point. Much more recently it was realized that the resulting states can have topological 
properties \cite{Volovik_2003, Armitage_Mele_Vishwanath_2018} that are robust agains small 
perturbations.\cite{Wan_et_al_2011, Burkov_Balents_2011, Burkov_Hook_Balents_2011} In three-dimensional systems the crossing point is generically
gapped out, unless the gap is fine tuned to zero. If the chemical potential lies in the gap, then the system will be an insulator that may have nontrivial
topological properties.\cite{Zhang_et_al_2009, Liu_et_al_2010} 
More generally, the chemical potential can lie within the conduction band. In this case the system is a true metal with a finite-size
Fermi surface. However, the underlying crossing point, whether or not it is gapped out, still leads to properties that are very different from those
of ordinary metals and independent of whether or not the material has nontrivial topological properties. In all of these cases the single-electron 
Hamiltonian in the vicinity of the crossing point is reminiscent of a Dirac Hamiltonian,
massless in the case of a gapless system, or massive in the case of a gapped one. In this paper we will consider the case of a generic chemical
potential, which makes the system a metal that we will refer to as a Dirac metal. A magnetic field, or a homogeneous magnetization, lifts the
degeneracy of the bands and separates the crossing points in momentum space. The single-particle spectrum then is reminiscent of the one
described by the Weyl equation.

Recently, we have considered electron-electron correlation effects in a such defined Dirac metal.\cite{Kirkpatrick_Belitz_2019a} In particular, we
have calculated the spin susceptibility $\chis$ at zero temperature ($T=0$) and have found it to be a nonanalytic function of an external magnetic field $h$. 
In a generic spatial dimension $d$ the leading nonanalytic contribution is proportional to $h^{d-1}$, and for $d=3$ it is $h^2\ln h$. This is a result of soft
or massless excitations in the underlying Dirac Fermi liquid that are rendered massive by a magnetic field. While the resulting nonanalyticity has the
same functional form as in an ordinary or Landau Fermi liquid,\cite{Belitz_Kirkpatrick_Vojta_1997, Betouras_Efremov_Chubukov_2005} this result
came as a surprise since the spin-orbit interaction gives a mass to the soft modes that are operative in its absence. However, it turns out that the
chirality degree of freedom in a Dirac metal leads to a new class of soft modes that have the same effect.

It is the chirality degree of freedom that makes the conduction-electron system in a Dirac metal form a type of Fermi liquid that is
qualitatively different from an ordinary or Landau Fermi liquid, and we refer to it as a Dirac Fermi liquid.
By contrast, generalizations of the original Landau Fermi-liquid theory to include a spin-orbit interaction \cite{Fujita_Quader_1987, Ashrafi_Rashba_Maslov_2013} 
still describe a Landau Fermi liquid in our nomenclature, as they do not contain the new class or soft modes that are crucial for our
purposes. Both the Landau and the Dirac Fermi liquid are true Fermi liquids in the sense that they have a finite Fermi surface, well-defined
quasiparticles, and the excitations in the interacting system are adiabatically connected to those of the underlying Fermi gas.

In ordinary (or Landau) metals it is known that the same soft modes that lead to the nonanalyticity in $\chis$ have a profound influence on the quantum
phase transition from a paramagnetic metal to a ferromagnetic one: They make the quantum phase transition in clean metals generically
first order.\cite{Belitz_Kirkpatrick_Vojta_1997, Belitz_Kirkpatrick_Vojta_1999, Brando_et_al_2016a} A nonzero temperature ($T>0$) gives
the soft modes a mass, which leads to a tricritical point in the phase diagram and to tricritical wings upon the application of a magnetic
field.\cite{Belitz_Kirkpatrick_Rollbuehler_2005} Numerous experiments have confirmed these predictions.\cite{Brando_et_al_2016a}
It further has been predicted that the quantum phase transition is first order in ferrimagnets and canted ferromagnets,\cite{Kirkpatrick_Belitz_2012b}
as well as in magnetic nematics.\cite{Kirkpatrick_Belitz_2011}

These observations raise the question whether the chiral soft modes in Dirac materials also lead to a first-order quantum phase transition in
ferromagnetic and related materials. This problem is of particular interest since some Dirac/Weyl materials are known or suspected to be 
ferromagnetic.\cite{Wan_et_al_2011, Xu_et_al_2011, Kubler_Felser_2016, Wang_et_al_2016b, Chang_et_al_2016, Liu_et_al_2018}
We will show that generically the quantum phase transition is indeed first order in Dirac metals, as it is in ordinary metals; however, systems
in which the gap vanishes due to a crystal symmetry may provide an exception. 

This paper is organized as follows. In Sec.~\ref{sec:II} we introduce the model, in Sec.~\ref{sec:III} we discuss the soft modes, and in Sec.~\ref{sec:IV}
we calculate the spin susceptibility. 
In Sec.~\ref{sec:V} 
we show how the nonanalytic behavior of $\chis$ leads to a first order quantum phase transition in ferromagnets, ferrimagnets, and magnetic
nematics. We conclude in Sec.~\ref{sec:VI} with a discussion of our results. Some of the results in Sec.~\ref{subsec:V.A} have been
reported before in a brief communication.\cite{Kirkpatrick_Belitz_2019b}
\section{Model}
\label{sec:II}

In this section we consider the same model as in Ref.~\onlinecite{Kirkpatrick_Belitz_2019a}, but give a more comprehensive discussion:
Rather than focusing entirely on gapless Dirac systems, we also consider the case where the gap is large (in a sense to be specified).
As we will see, this is important for determining which correlations can contribute to the nonanalytic behavior of the spin susceptibility
and thus influence the nature of the magnetic quantum phase transitions.

\subsection{Model for a chiral Fermi gas}
\label{subsubsec:II.A}

\subsubsection{Single-particle Hamiltonian}

\begin{figure*}[t]
\includegraphics[width=17cm]{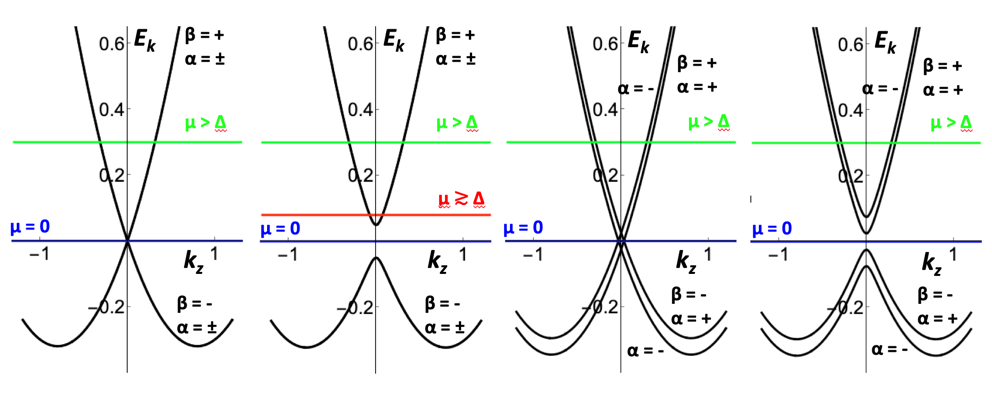}
\caption{Single-particle spectra for $v=0.8$ and (from left to right) $\Delta = h = 0$; $\Delta=0.05$, $h=0$; $\Delta = 0$, $h=0.025$;
$\Delta=0.05$, $h=0.025$ in atomic units. The large value of $v$ emphasizes the cone structure of the spectrum. For a zero chemical
potential $\mu$ the system is a semimetal for $\Delta=0$ and an insulator for $\Delta>0$, for a sufficiently large chemical potential it
is a metal in all cases. Note that the lower-cone branches ($\beta=-1$) do not contribute to the Fermi surface of the metal.
}
\label{fig:1}
\end{figure*}
We consider systems where the spin-orbit interaction leads to a linear band crossing via a term proportional to ${\bm k}\cdot{\bm\sigma}$
in the single-particle Hamiltonian, with ${\bm\sigma} = (\sigma_1,\sigma_2,\sigma_3)$ the spin Pauli matrices.\cite{Abrikosov_Beneslavskii_1970} 
Such a term is invariant under time reversal, but not under spatial inversions. If the system is invariant under the latter, the Hamiltonian therefore 
must contain both left-handed and right-handed electrons.\cite{Liu_et_al_2010, Burkov_2015} The physical origin of this chirality degree of
freedom is the fact that the crossing bands have different parities.
It is convenient to encode it via a second set of Pauli matrices ${\bm\pi} = (\pi_1,\pi_2,\pi_3)$. With
$\sigma_0 = \pi_0$ the $2\times 2$ unit matrix, the most general single-particle Hamiltonian that  is invariant under both time reversal
and spatial inversions then can be written\cite{Zhang_et_al_2009, Liu_et_al_2010}
\bea
H_0 &=& (\epsilon_{\bm k} - \mu)(\pi_0\otimes\sigma_0) + v(\pi_3\otimes{\bm\sigma})\cdot{\bm k} +\Delta(\pi_1\otimes\sigma_0) 
\nonumber\\
&&  - h(\pi_0\otimes\sigma_3) \ .
\label{eq:2.1}
\eea
The first term is an ordinary band Hamiltonian with a single-particle energy $\epsilon_{\bm k} = \epsilon_{-{\bm k}}$ that is quadratic for small ${\bm k}$.
For simplicity, will take $\epsilon_{\bm k} = {\bm k}^2/2m$ with an effective mass $m$. $\mu$ is the chemical potential. The second term is the spin-orbit 
coupling introduced above, with a coupling constant $v$ that dimensionally is a velocity. The third term, with coupling constant $\Delta$,
also respects both time reversal and spatial inversion.\cite{Delta_footnote} It mixes left- and right-handed electrons in a symmetric way and thus breaks 
a gauge symmetry that expresses the conservation of the number of electrons with a given handedness or chirality. The last term is a Zeeman term with
a magnetic field ${\bm h} = (0,0,h)$ in the 3-direction that breaks time reversal.
The chirality degree of freedom, encoded in the Pauli matrices ${\bm\pi}$, is crucial for the soft-mode structure of the system, as we
will see below. In chirality space, $v$ acts as a longitudinal field, while $\Delta$ acts as a transverse field. For a vanishing spin-orbit interaction, 
$v=\Delta=0$, Eq.~(\ref{eq:2.1}) reduces to the Hamiltonian for an ordinary (or Landau) Fermi gas with single-particle energy $\epsilon_{\bm k}$. 

\subsubsection{Single-particle spectrum}
\label{subsubsec:II.A.2}

%
The single-particle spectrum is readily obtained by finding the eigenvalues $\lambda_{\bm k}$ of the $4\times 4$ Hamiltonian $H_0$. 
For the four eigenvalues one finds
\bse
\label{eqs:2.2}
\be
\lambda_{\bm k}^{\alpha\beta} = \xi_{\bm k} + \beta \vert v{\bm k}_{\Delta} - \alpha {\bm h}\vert\ ,
\label{eq:2.2a}
\ee
where $\xi_{\bm k} = \epsilon_{\bm k} - \mu$, and $\alpha,\beta = \pm 1$. The four branches of the
single-particle energy $E_{\bm k} = \lambda_{\bm k} + \mu$ are
\be
E_{\bm k}^{\alpha\beta} = \epsilon_{\bm k} + \beta \vert v{\bm k}_{\Delta} - \alpha {\bm h}\vert\ .
\label{eq:2.2b}
\ee
\ese
We will refer to $\beta$ as the cone index, and to $\alpha$ as the chirality index. In Eqs.~(\ref{eqs:2.2}) we have defined
\be
{\bm k}_{\Delta} = \left(k_x, k_y, s_{\bm k}\sqrt{k_z^2 + \Delta^2/v^2}\right)\ .
\label{eq:2.3}
\ee
For $s_{\bm k}$ one can choose either $s_{\bm k} = \sgn(k_z)$, or $s_{\bm k} = 1$. The functional form of the
spectrum is independent of this choice, the two choices just amount to a relabeling of the branches of the
spectrum in a nonzero magnetic field. For explicit calculations in the limit $\Delta = 0$ the choice 
 $s_{\bm k} = \sgn(k_z)$ is more convenient, as it results in
 \bse
 \label{eqs:2.4}
 \be
 {\bm k}_{\Delta} = {\bm k}\qquad (\Delta = 0)\ .
 \label{eq:2.4a}
 \ee
 In the limit $\Delta \gg v\kF$, with $\kF$ the Fermi wave number, the choice $s_{\bm k} = 1$ is more convenient,
 which results in
 \be
 {\bm k}_{\Delta} \approx (k_x,k_y,\Delta/v)\qquad (\Delta \gg v\kF)\ .
 \label{eq:2.4b}
 \ee
 \ese

For a vanishing spin-orbit interaction, $v=\Delta=0$, we have $v{\bm k}_{\Delta} = 0$, the spectrum is two-fold
degenerate in the chirality index, and $\beta$ reduces to the spin projection $\sigma = \pm$. For $v\neq 0$ the
spectrum is still two-fold degenerate in zero field, but a magnetic field splits this degeneracy. 

In order to illustrate the shapes of the spectrum for different ranges of parameter values, let us introduce an atomic-scale
momentum $p_0$ (on the order of an inverse lattice spacing), 
velocity $v_0 = p_0/2m$, and energy $E_0 = p_0^2/2m$. We then measure $E_{\bm k}$, $\Delta$, and $h$ in
units of $E_0$, 
$v$ in units of $v_0$, and ${\bm k}$ in units of $k_0$. 
\begin{figure*}[t]
\includegraphics[width=17cm]{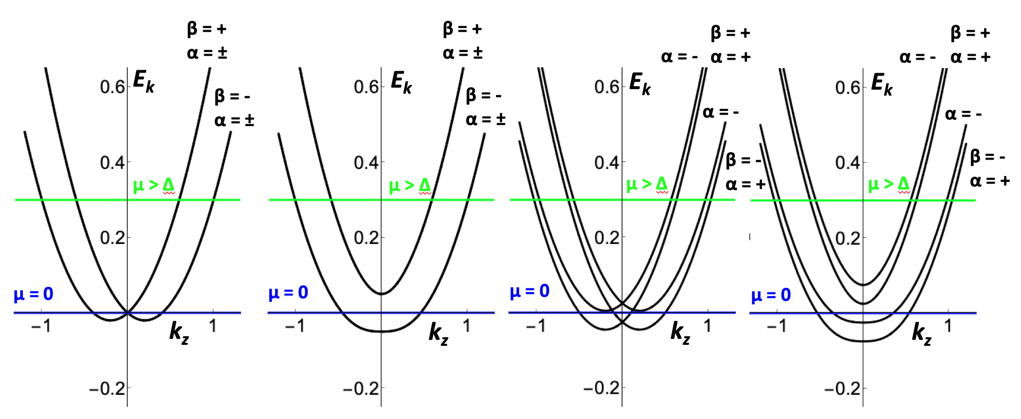}
\caption{Same as Fig.~\ref{fig:1}, but for $v=0.2$. Both the upper and the lower-cone branches contribute to the Fermi surface.}
\label{fig:2}
\end{figure*}

Figure~\ref{fig:1} shows the spectrum for $v$ on the order of the atomic velocity scale, $v\alt v_0$. 
The linear term then dominates, and the spectrum has a characteristic cone structure. 
For $h=0$, each cone is two-fold degenerate with respect to the chirality index $\alpha$. This makes the spectrum reminiscent
of a massless (for $\Delta=0$) or massive (for $\Delta>0$) Dirac equation. For a vanishing chemical potential
($\mu=0$) the system is a Dirac semimetal or insulator, respectively. For $\mu > \Delta$ it is a true metal that we
will refer to as a Dirac metal. The lower cone ($\beta = -1$) does not turn up before the
edge of the Brillouin zone is reached, and only the upper cone ($\beta = +1$) contributes to the Fermi surface,
which is defined by
\be
\mu = E_{\bm k}^{\alpha\beta}\bigr\vert_{{\bm k}\in\text{FS}}
\label{eq:2.5}
\ee
To determine the Fermi wave number, we distinguish between two cases. Let $\Delta \ll v\kF$. Then
\bse
\label{eqs:2.6}
\be
\kF = \begin{cases} \sqrt{2 m \mu} \quad & \text{if} \quad m v^2 \ll \mu \\
                                \mu/v                        & \text{if} \quad m v^2 \gg \mu
         \end{cases}
         \qquad (\Delta \ll v\kF < \mu)\ .
\label{eq:2.6a}
\ee
Now let $\Delta \gg v\kF$. Then
\be
\kF = \sqrt{2m(\mu - \Delta)}\qquad (\mu > \Delta \gg v\kF)\ .
\label{eq:2.6b}
\ee
\ese
Note that this case is realizable provided $\mu \agt \Delta$, even though $v$ is on the order of $v_0$.
This Fermi surface is still degenerate in the chirality index.  A magnetic field lifts the degeneracy in $\alpha$ 
and separates the cones in $k$-space, and the system becomes a Weyl semimetal or insulator for $\mu=0$, and a Weyl 
metal for $\mu > \Delta$. All of this is demonstrated in Fig.~\ref{fig:1}. We also note that in systems that are 
not invariant under spatial inversion, or if time reversal is broken by effects other than a magnetic field, there are additional possibilities, 
in addition to the Dirac and Weyl cases, that we will not discuss, see, e.g., Ref.~\onlinecite{Mitchell_Fritz_2015}.

Figure~\ref{fig:2} shows the spectrum for a value of $v$ that is much smaller, but still significant on the atomic scale. The cone
structure is now visible only for small $k$, the lower cone turns up for $k \ll p_0$, and both cones contribute to the
Fermi surface if $\mu > \Delta$. Although the structure of the spectrum for small $k$ is rather different in the two
cases, the only qualitative difference near the Fermi surface in the metallic case ($\mu > \Delta$) is the number
of contributing cone branches. Indeed, we will show that the soft-mode spectrum is the same irrespective of the
value of $v$ (as long as $v$ is significant in a sense to be made explicit), and it makes sense to refer to all of
these systems as Dirac or Weyl metals, respectively. For $v = \Delta = h = 0$ the spectrum reduces to the four-fold
degenerate parabolic band of an ordinary nearly-free electron gas with an additional degree of freedom. An
observation that will be very important for our purposes is that $v>0$ at $h=0$ splits the band in a way that is very similar
to the splitting by $h>0$ at $v=0$: For $\Delta \ll \sqrt{m v^2 \mu} \ll \mu$ we have Fermi wave numbers
\be
\kF^{\beta} = \sqrt{2 m \mu} - \beta m v + O(v^2) \quad (\Delta \ll \sqrt{m v^2 \mu} \ll \mu)\ .
\label{eq:2.7}
\ee
Each of these two Fermi surfaces is still two-fold degenerate, and $h>0$ lifts that degeneracy. All of these
features are demonstrated in Fig.~\ref{fig:2}.

\subsubsection{Green function}
\label{subsubsec:II.A.3}


The single-particle Green function is defined as
\be
G_k = \left[i\omega_n(\pi_0\otimes\sigma_0) - H_0\right]^{-1}\ ,
\label{eq:2.8}
\ee
where $k = (i\omega_n,{\bm k})$ is a 4-vector comprising a fermionic Matsubara frequency $\omega_n$ and a
wave vector ${\bm k}$. An exact expression for $G_k$ in terms of the quasiparticle resonances
\be
F_k^{\alpha\beta} = \frac{1}{i\omega_n - \lambda_{\bm k}^{\alpha\beta}} = \frac{1}{i\omega_n - \xi_{\bm k} - \beta \vert v{\bm k}_{\Delta} - \alpha{\bm h}\vert}
\label{eq:2.9}
\ee
is given in Appendix~\ref{app:A}. While exact, it is not suitable for explicit calculations, and it is desirable
to perform a partial fraction decomposition to write $G_k$ in the form
\be
G_k = \sum_{\alpha,\beta} F_k^{\alpha\beta}\,M^{\alpha\beta}({\bm k})
\label{eq:2.10}
\ee
with spin-chirality matrices $M^{\alpha\beta}$. In general, the latter are very complicated. However, for our
purposes we do not need the complete expressions. The leading nonanalytic $h$-dependence of the spin
susceptibility arises from the $h$-dependence of the denominator in $F_k^{\alpha\beta}$, which cuts off
the singularity of the quasiparticle resonance. We can therefore evaluate the numerator  in the limit $h\ll v\kF$.
(Note that this precludes taking the limit $v\to 0$, see Sec.~\ref{sec:III}.)
Additional simplifications occur in the limits $\Delta = 0$ and $\Delta \gg v\kF$. Using Eq.~(\ref{eq:2.4a})
and (\ref{eq:2.4b}), respectively, in these two cases we find
\bse
\label{eqs:2.11}
\be
M_{\Delta=0}^{\alpha\beta}({\bm k}) \approx  \frac{1}{4}\, (\pi_0 + \alpha\pi_3)\otimes (\sigma_0 + \alpha\beta \hat{\bm k}\cdot{\bm\sigma})\ ,
\label{eq:2.11a}
\ee
which depends only on the unit vector $\hat{\bm k} = {\bm k}/\vert{\bm k}\vert$, and
\be
M_{\Delta\gg v\kF}^{\alpha\beta} \approx \frac{1}{4}\left(\pi_0 + \beta \pi_1\right) \otimes \left(\sigma_0 \pm \alpha\beta\sigma_3\right)\ ,
\label{eq:2.11b}
\ee
\ese
which is independent of ${\bm k}$. $F_k^{\alpha\beta}$ in the two cases is given by Eq.~(\ref{eq:2.10}) with ${\bm k}_{\Delta}$
from Eq.~(\ref{eq:2.4a}) and (\ref{eq:2.4b}), respectively. Equation (\ref{eq:2.11a}) together with Eqs.~(\ref{eq:2.10}), (\ref{eq:2.9})
is equivalent to the expression for the Green function in Ref.~\onlinecite{Kirkpatrick_Belitz_2019a}. Note that $G_k$ for
$\Delta \gg v\kF$ can be written in the form of Eq.~(\ref{eq:2.10}) plus (\ref{eq:2.11b}) {\em only} with the choice $s_{\bm k} = 1$
in Eq.~(\ref{eq:2.3}).

We stress that Eqs.~(\ref{eqs:2.11}) are valid only for calculating the leading nonanalytic $h$-dependence of the spin susceptibility in these two limits,
and for $v\neq 0$. In particular, they do no longer allow for taking the Landau limit $v=\Delta=0$. The latter can of course be recovered from
the exact expression for the Green function in Appendix~\ref{app:A} and can be written in a form analogous to Eq.~(\ref{eq:2.10}). We find
\bse
\label{eqs:2.12}
\be
G_k = \sum_{\sigma=\pm} G_k^{\sigma}\,M^{\sigma}(\hat{\bm h}) \qquad (v=\Delta=0)\ ,
\label{eq:2.12a}
\ee
where
\be
G_k^{\sigma} = F_k^{\alpha\sigma} = \frac{1}{i\omega_n - \xi_{\bm k} + \sigma h}\ ,
\label{eq:2.12b}
\ee
which now is independent of $\alpha$, and
\be
M^{\sigma}(\hat{\bm h}) = \frac{1}{2}\,\pi_0\otimes\left(\sigma_0 - \sigma\hat{\bm h}\cdot{\bm\sigma}\right)\ .
\label{eq:2.12c}
\ee
\ese

\subsubsection{Action for noninteracting electrons}
\label{subsubsec:II.A.4}

The electrons of our Dirac Fermi gas are described in terms of fermionic fields $\bar\psi_{\sigma}^{\pi}(k)$ and $\psi_{\sigma}^{\pi}(k)$
that carry a spin index $\sigma = \uparrow,\downarrow \equiv \pm$ and a chirality index $\pi = \pm$. Introducing spinors
$\psi = (\psi_{\uparrow}^+, \psi_{\downarrow}^+, \psi_{\uparrow}^-, \psi_{\downarrow}^-)$ and a scalar product 
$(\bar\psi,\psi) = \sum_{\sigma,\pi} \bar\psi_{\sigma}^{\pi} \psi_{\sigma}^{\pi}$ we can write the action of the noninteracting
fermion system governed by the Hamiltonian $H_0$ in terms of the inverse Green function,
\be
S_0 = \sum_k \left(\bar\psi(k), \left[ i\omega_n (\pi_0 \otimes \sigma_0) - H_0 \right] \psi(k) \right)\ .
\label{eq:2.13}
\ee

\subsection{Electron-electron interaction}
\label{subsec:II.B}

\begin{figure}[b]
\includegraphics[width=8.5cm]{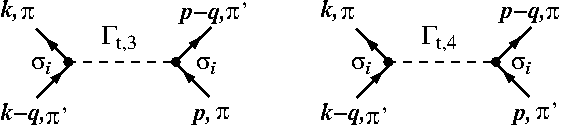}
\caption{Spin-triplet interaction amplitudes that mix chiralities.}
\label{fig:3}
\end{figure}

The noninteracting action, Eq.~(\ref{eq:2.13}), needs to be supplemented by all four-fermion interaction terms that
respect the same symmetries as $H_0$. We are interested in true metals, hence screening works and we can
consider interaction amplitudes that are point-like in space and time. The noninteracting action is invariant under
simultaneous rotations in spin and momentum space, in addition to spatial inversions and time reversal. There are
eight interaction terms that respect these requirements:

\begin{widetext}
\bea
S_{\text{int}} &=& \frac{-T}{2V} \sum_{q}{}^{'}  \sum_{k,p} \sum_{i=0}^3  \Bigl[ \tilde\Gamma_{\text{s,i}} 
     \left(\bar\psi(k),(\sigma_0\otimes\pi_i)\psi(k-q)\right)\left(\bar\psi(p-q),(\sigma_0\otimes\pi_i)\psi(p)\right)
\nonumber\\
   && \hskip 75pt -\tilde\Gamma_{\text{t,i}} 
     \left(\bar\psi(k),(\bm\sigma\otimes\pi_i)\psi(k-q)\right)\cdot\left(\bar\psi(p-q),(\bm\sigma\otimes\pi_i)\psi(p)\right)
                 \Bigr]
 \label{eq:2.14}
  \eea
 with interaction amplitudes $\tilde\Gamma_{\text{s},i}$ and $\tilde\Gamma_{\text{t},i}$ ($i=0,1,2,3$) in the
 spin-singlet and spin-triplet channels, respectively. Note that the rotational invariance in spin space
 ensures that all three spin-triplet amplitudes for a given chirality channel $i$ are equal, whereas there is no
 equivalent requirement in chirality space.\cite{crystal_field_footnote} The prime on the sum over $q$
 indicates a restriction to $\vert{\bm q}\vert < \Lambda$, with $\Lambda \ll \kF$ a momentum cutoff.
 This is necessary in order to avoid double counting, as explained in Ref.~\onlinecite{Kirkpatrick_Belitz_2019a}.
 With this choice of interaction amplitudes, the $\tilde\Gamma$ all carry a small, or hydrodynamic,
 wave vector ${\bm q}$. For illustrative purposes, and to make contact with Ref.~\onlinecite{Kirkpatrick_Belitz_2019a}, 
 we also give the interaction terms with the chirality index written explicitly: 
\begin{align}
\nonumber\\
S_{\text{int}} &= \frac{-T}{2V} \sum_{q}{}^{'}  \sum_{k,p} 
\nonumber\\
&\hskip -15pt \times \Bigl[ \Gamma_{\text{s,1}} \sum_{\pi}
     \left(\bar\psi^{\pi}(k),\sigma_0\psi^{\pi}(k-q)\right)\left(\bar\psi^{\pi}(p-q),\sigma_0\psi^{\pi}(p)\right)
 +\Gamma_{\text{s,2}} \sum_{\pi\neq\pi'}
     \left(\bar\psi^{\pi}(k),\sigma_0\psi^{\pi}(k-q)\right)\left(\bar\psi^{\pi'}(p-q),\sigma_0\psi^{\pi'}(p)\right)
\nonumber\\
&  \hskip -6pt +\Gamma_{\text{s,3}} \sum_{\pi\neq\pi'}
     \left(\bar\psi^{\pi}(k),\sigma_0\psi^{\pi'}(k-q)\right)\left(\bar\psi^{\pi'}(p-q),\sigma_0\psi^{\pi}(p)\right)
   +\Gamma_{\text{s,4}} \sum_{\pi\neq\pi'}
     \left(\bar\psi^{\pi}(k),\sigma_0\psi^{\pi'}(k-q)\right)\left(\bar\psi^{\pi}(p-q),\sigma_0\psi^{\pi'}(p)\right)
\nonumber\\
& \hskip -6pt -\Gamma_{\text{t,1}} \sum_{\pi}
     \left(\bar\psi^{\pi}(k),\bm\sigma\psi^{\pi}(k-q)\right)\cdot\left(\bar\psi^{\pi}(p-q),\bm\sigma\psi^{\pi}(p)\right)
-\Gamma_{\text{t,2}} \sum_{\pi\neq\pi'}
     \left(\bar\psi^{\pi}(k)\bm\sigma\psi^{\pi}(k-q)\right)\cdot\left(\bar\psi^{\pi'}(p-q)\bm\sigma\psi^{\pi'}(p)\right)
\nonumber\\
&  \hskip -6pt   -\Gamma_{\text{t,3}} \sum_{\pi\neq\pi'}
     \left(\bar\psi^{\pi}(k),\bm\sigma\psi^{\pi'}(k-q)\right)\cdot\left(\bar\psi^{\pi'}(p-q),\bm\sigma\psi^{\pi}(p)\right) 
 - \Gamma_{{\text t},4} \sum_{\pi\neq\pi'} \left(\bar\psi^{\pi}(k),{\bm\sigma}\psi^{\pi'}(k-q)\right)\cdot  \left(\bar\psi^{\pi}(p-q),{\bm\sigma}\psi^{\pi'}(p)\right) 
                 \Bigr].
\tag{$2.14'$}      
\end{align}
\end{widetext}
Here $\psi^{\pi} = (\psi_{\uparrow}^{\pi},\psi_{\downarrow}^{\pi})$ and $\bar\psi^{\pi} = (\bar\psi_{\uparrow}^{\pi},\bar\psi_{\downarrow}^{\pi})$
are two-component spinors, and the $\Gamma$ are linear combinations of the $\tilde\Gamma$ : $\Gamma_{\text{s},1} = \tilde\Gamma_{\text{s},0} + \tilde\Gamma_{\text{s},3}$, etc.
Note that $\Gamma_{\text{s},4}$ and $\Gamma_{\text{t},4}$ break the same gauge symmetry as the $\Delta$ term in $H_0$.
They were not considered in Ref.~\onlinecite{Kirkpatrick_Belitz_2019a}, which focused on the same model with $\Delta=0$.
For our purposes we will be particularly interested in spin-triplet interactions that mix chiralities, i.e., in the amplitudes $\Gamma_{\text{t},3}$ and
$\Gamma_{\text{t},4}$. They are graphically represented in Fig.~\ref{fig:3}.

The Eqs.~(\ref{eq:2.13}) and (\ref{eq:2.14}) or (2.14') completely specify our model for a Dirac metal:
\be
S_{\text{DM}} = S_0 + S_{\text{int}}\ .
\label{eq:2.15}
\ee

\section{Soft modes in Landau and Dirac metals}
\label{sec:III}

Soft modes, i.e., correlation functions that diverge in the limit of vanishing wave vector and frequency, are
responsible for all nonanalytic behavior of observables. In a Fermi liquid, whether of Landau or of Dirac
type, single-particle excitations are well known to be soft. This property can immediately be seen in the
Green function, i.e., the two-fermion correlation function, which diverges at zero frequency with the wave 
vector on the Fermi surface. Less well known is a class of soft two-particle excitations, or four-fermion correlation 
functions, although their consequences have been known for a long time. To explain the nature of these
soft modes, which are the ones relevant for our purposes, it is illustrative to first discuss their
manifestations in a Landau Fermi liquid.

\subsection{Soft modes in a Landau Fermi liquid}
\label{subsec:III.A}

Consider the Green function for a Landau Fermi gas as written in Eqs.~(\ref{eqs:2.12}). An explicit
calculation easily proves Velicky's Ward identity,\cite{Velicky_1969}
\bea
G_{i\omega_{n_1},{\bm k}+{\bm q}/2}^{\sigma_1} \, G_{i\omega_{n_2},{\bm k}-{\bm q}/2}^{\sigma_2} &=&
\nonumber\\
&& \hskip -80pt \frac{-\left(G_{i\omega_{n_1},{\bm k}+{\bm q}/2}^{\sigma_1} - G_{i\omega_{n_2},{\bm k}-{\bm q}/2}^{\sigma_2}\right)}
                                 {i\Omega_{n_1-n_2} - {\bm k}\cdot{\bm q}/m + (\sigma_1-\sigma_2)h}
\label{eq:3.1}
\eea
where $\Omega_{n_1-n_2} = \omega_{n_1} - \omega_{n_2}$ is a bosonic Matsubara frequency. It relates a
four-fermion correlation function on the left-hand side (which factorizes since we are dealing with noninteracting
electrons) to the difference of two two-fermion correlations on the right-hand side. The salient point is the
structure of the right-hand side: If $h=0$, or $\sigma_1 = \sigma_2$, then the denominator vanishes in the
limit $\omega_{n_1}, \omega_{n_2}, {\bm q} \to 0$. By contrast, the numerator vanishes only if $\omega_{n_1}$
and $\omega_{n_2}$ have the same sign, whereas it is nonzero if these two frequencies have opposite signs,
due to the cut of the Green function on the real axis. Four-fermion correlations of the type represented by the
left-hand side with Matsubara frequencies on opposite sides of the real axis thus are soft modes with a
ballistic frequency-momentum relation, where the frequency scales as the wave number, $\Omega \sim {\bm q}$.
Consistent with that, the denominator on the right-hand side has the structure of a Boltzmann equation in the
absence of a collision operator, namely, a time derivative plus a streaming term. For spin projections that are
not the same, $\sigma_1 \neq \sigma_2$, a magnetic field $h$ gives the soft mode a mass that also scales
as the wave number. Finally, temperature scales linearly with frequency, so frequency, wave number,
magnetic field, and temperature all scale the same way, $\Omega \sim {\bm q} \sim h \sim T$. 

This deceptively simple structure is very robust. It also holds, with the ballistic dynamics replaced by diffusive
ones, in the presence of quenched disorder if one takes $G$ to be the unaveraged Green function and performs
a disorder average on either side of the identity.\cite{Velicky_1969, Wegner_1979} Remarkably, the resulting
four-fermion correlations, often referred to as ``diffusons'', are soft even though the single-particle excitations
are now massive due to the quenched disorder. They play an important role in the theory of Anderson
localization\cite{Lee_Ramakrishnan_1985, Altshuler_Aronov_1984} and universal conductance
fluctuations.\cite{Lee_Stone_1985, Akkermans_Montambaux_2011} Wegner has shown that the diffusons
are properly interpreted as the Goldstone modes of a spontaneously broken symmetry that can be
formulated as a rotational symmetry in frequency space, or the symmetry between retarded and
advanced degrees of freedom.\cite{Wegner_1979, Schaefer_Wegner_1980} The same interpretation holds
for the ballistic soft modes in clean systems.\cite{Belitz_Kirkpatrick_2012a, Kirkpatrick_Belitz_2019a} This
explains their robustness, and strongly suggests that they remain soft in the presence of an electron-electron
interaction, since the latter cannot change the analytic structure reflected in the difference between
retarded and advanced degrees of freedom. Indeed, Eq.~(\ref{eq:3.1}) remains valid, with a slightly more complicated frequency structure,
and a renormalized electron mass $m$, if the left-hand side is replaced by an appropriate four-fermion
correlation that factorizes into the two Green functions shown in the noninteracting limit. This has been
shown with and without quenched disorder in Refs.~\onlinecite{Belitz_Kirkpatrick_1997} and
\onlinecite{Belitz_Kirkpatrick_2012a}, respectively. At nonzero temperature the interaction leads to
a dephasing rate that gives the soft modes a mass. However, this rate goes as $T^{d-1}$ in $d$
spatial dimensions, so this effect is subleading compared to the $\Omega \sim T$ scaling in all
dimensions that support a Fermi liquid.

It is illustrative to consider wave-vector convolutions of the form
\bse
\label{eqs:3.2}
\bea
\varphi^{\sigma_1\sigma_2}({\bm q},i\Omega_n) &\equiv& \frac{1}{V}\sum_{\bm k} G_k^{\sigma_1}\,G_{k-q}^{\sigma_2}
\label{eq:3.2a}\\
&& \hskip -60pt =  \int \frac{d\Omega_{\bm k}}{4\pi}\,\frac{2\pi i \NF\,\sgn(\omega_m)\,\Theta(-\omega_m(\omega_m - \Omega_n))}{i\Omega_n - \vF {\hat{\bm k}}\cdot{\bm q} 
       + (\sigma_1 - \sigma_2)h} \ ,
       \nonumber\\
\label{eq:3.2b}
\eea
\ese
as these commonly appear in the calculation of observables. Here $q = (i\Omega_n,{\bm q})$, $\NF$ is the density of states per spin at the 
Fermi level, and $\vF$ is the Fermi velocity. The momentum integral in Eq.~(\ref{eq:3.2a})
has been performed in the well-known approximation that is valid for the leading behavior in the limit of small frequency and
wave number. It consists of replacing the radial part of the integral by an integration
over all real values of $\xi_{\bm k}$, and we will refer to it as the AGD approximation after Ref.~\onlinecite{Abrikosov_Gorkov_Dzyaloshinski_1963}.
The convolution $\varphi$, but with Green functions that contain an elastic scattering rate, is a crucial element of the
diffuson excitation in disordered systems mentioned above.\cite{Vollhardt_Woelfle_1980}

To summarize, a clean Landau Fermi liquid at $T=0$ contains soft two-particle excitations that are ballistic in nature, scale as $1/q$ (where $q$
can be a frequency or a wave number), and
acquire a mass at $T>0$. Obviously, momentum convolutions of $n>2$ Green functions will scale as $1/q^{n-1}$ provided the $n$
frequencies carried by the Green functions do not all have the same sign.

\subsection{Soft modes in a Dirac Fermi liquid}
\label{subsec:III.B}


Arguments that are structurally identical to those in the previous subsection yield information about the soft-mode
structure of a Dirac Fermi liquid, but the results are different in important ways. Consider Eqs.~(\ref{eqs:3.2})
with $G_k^{\sigma}$ replaced by $F_k^{\alpha\beta}$ from Eq.~(\ref{eq:2.9}). Instead of Eq.~(\ref{eq:3.1}) we find
\begin{widetext}
\bea
F_{i\omega_{n_1},{\bm k}+{\bm q}/2}^{\alpha_1\beta_1}\,F_{i\omega_{n_2},{\bm k}-{\bm q}/2}^{\alpha_2\beta_2} =
    \frac{-\left(F_{i\omega_{n_1},{\bm k}+{\bm q}/2}^{\alpha_1\beta_1} - F_{i\omega_{n_2},{\bm k}-{\bm q}/2}^{\alpha_2\beta_2}\right)}
                                 {i\Omega_{n_1 - n_2} - {\bm k}\cdot{\bm q}/m + \beta_1\vert v({\bm k}+{\bm q}/2)_{\Delta} - \alpha_1{\bm h}\vert
                                                                                                          - \beta_2\vert v({\bm k}-{\bm q}/2)_{\Delta} - \alpha_2{\bm h}\vert}
\label{eq:3.3}
\eea     
For the convolutions analogous to Eqs.~(\ref{eqs:3.2}) this yields
\bse
\label{eqs:3.4}
\bea
\phi^{\beta_1\beta_2}_{\alpha_1\alpha_2}({\bm q},i\Omega_n) &\equiv&  \frac{1}{V}\sum_{\bm k} F_k^{\alpha_1\beta_1}\,F_{k-q}^{\alpha_2\beta_2}
\label{eq:3.4a}\\
&=& 2\pi i \NF\,\sgn(\omega_m)\,\Theta(-\omega_m(\omega_m - \Omega_n)) \int \frac{d\Omega_{\bm k}}{4\pi}\,\frac{1}{N({\bm q},i\Omega_n;{\hat{\bm k}},i\omega_m)} 
\label{eq:3.4b}
\eea
In the limits $\Delta=0$ and $\Delta\gg v\kF$, respectively, the denominator is given by
\be
N({\bm q},i\Omega_n;{\hat{\bm k}},i\omega_m) = \begin{cases} i\Omega_n - \vF{\hat{\bm k}}\cdot{\bm q} + \beta_2\vert v(\kF \hat{\bm k} - {\bm q}) - \alpha_2 {\bm h}\vert 
     - \beta_1\vert v\kF \hat{\bm k} - \alpha_1{\bm h}\vert \quad \text{for}\quad \Delta=0\\
i\Omega_n - \vF{\hat{\bm k}}\cdot{\bm q} + \beta_2\vert\Delta - \alpha_2 h\vert - \beta_1\vert\Delta - \alpha_1 h\vert \hskip 66pt \text{for}\quad \Delta \gg v\kF\ .
\end{cases}
\label{eq:3.4c}
\ee
\ese
\end{widetext}      
For $v = \Delta = 0$ the cone index $\beta$ reverts to the spin projection index $\sigma$ and we recover Eqs.~(\ref{eqs:3.2}). To discuss the Dirac case,
let us first consider $h=0$. Now the modes with $\beta_1 \neq \beta_2$ are massive, with the mass
determined by $v\kF$, or $\Delta$, or both. That is, the modes that are massless in a Landau Fermi liquid at $h=0$ and acquire a mass for $h\neq 0$,
see Eq.~(\ref{eq:3.2b}), are massive in a Dirac Fermi liquid. This was to be expected, since $v$ splits the Fermi surface of a Dirac metal in much the
same way as $h$ does in a Landau metal, see Fig.~\ref{fig:2}. However, the modes with $\beta_1 = \beta_2$ are soft, and in a magnetic field they acquire 
a mass provided $\alpha_1 \neq \alpha_2$. The chirality degree of freedom thus provides for a new class of soft modes that are cut off by a magnetic
field. This will be of crucial importance in what follows.

\subsection{Physical consequences of soft modes}
\label{subsec:III.C}      

It is obvious that soft modes that are cut off by an external field will result in the free energy being a nonanalytic
function of that field, and the same will be true for all derivatives of the free energy with respect to the field. This
has been discussed in some detail in Refs.~\onlinecite{Belitz_Kirkpatrick_Vojta_2002} and \onlinecite{Kirkpatrick_Belitz_2019a},
so here we give only a brief summary. The external field relevant for our purposes is the external magnetic field $h$, which
is conjugate to the spin density, so we need to distinguish soft modes that are given a mass by $h$ from those that are not.
In the nomenclature of Ref.~\onlinecite{Kirkpatrick_Belitz_2019a}, such modes are soft of the first kind and the second kind,
respectively, with respect to $h$. From Eq.~(\ref{eq:3.4c}) we see that, in a Dirac metal, the modes with $\beta_1 = \beta_2$
and $\alpha_1 \neq \alpha_2$ are of the first kind with respect to $h$, all others are either soft of the second kind, or massive. 
In this context we stress that there is an important conceptual difference between the mass provided by $h$ compared to the
mass provided by $v$ or $\Delta$ in Eqs.~(\ref{eqs:3.4}). $v$ and $\Delta$ are not considered tunable fields for our purposes,
and we are not interested in derivatives of the free energy with respect to them. Their effect is entirely to make certain modes
irrelevant for determining the hydrodynamic (i.e., long-wavelength/low-frequency) properties of the system. $h$, on the other
hand, is a tunable field and the second derivative of the free energy with respect to it gives the spin susceptibility. $h$
providing a mass for a soft mode thus results in a spin susceptibility being a nonanalytic function of $h$. This is interesting
in its own right and has important consequences for magnetic quantum phase transitions, as we will see.

\section{The spin susceptibility}
\label{sec:IV}

We now calculate the longitudinal spin susceptibility $\chis^{\text L}$ of a Dirac Fermi liquid, given by the 2-point correlation
function of the 3-component of the spin density. We will focus on the $h$-dependence and calculate the homogeneous,
static susceptibility
\bse
\label{eqs:4.1}
\be
\chis^{\text{L}}(h) = \frac{T}{V} \int dx\,dy\ \langle \delta n_{\text{s}}^3(x)\,\delta n_{\text{s}}^3(y) \rangle_{S_{\text{DM}}}\ .
\label{eq:4.1a}
\ee
Here $x = ({\bm x},\tau)$ is a 4-vector comprising the real-space position ${\bm x}$ and the imaginary-time variable
$\tau$, $n_{\text{s}}^3$ is the 3-component of the spin density,
\be
n_{\text{s}}^3(x) = \left(\bar\psi(x),(\pi_0\otimes\sigma_3) \psi(x)\right)\ ,
\label{eq:4.1b}
\ee
and
\be
\delta n_{\text{s}}^3(x) = n_{\text{s}}^3(x) - \langle n_{\text{s}}^3(x) \rangle_{S_{\text{DM}}}\ .
\label{eq:4.1c}
\ee
\ese
The averages are with respect to the Dirac-metal action $S_{\text{DM}}$ from Eq.~(\ref{eq:2.15}), which
includes the magnetic field via the Zeeman term in Eq.~(\ref{eq:2.1}). In Fourier space,
\bse
\label{eq:4.2}
\be
\chis^{\text L}(k,h) = \langle \delta n_{\text s}^3(k)\,\delta n_{\text s}^3(-k)\rangle_{S_{\text{DM}}}
\label{eq:4.2a}
\ee
with 
\bea
n_{\text s}^3(k) &=& \sum_p \left(\bar\psi(p),(\sigma_3\otimes\pi_0)\psi(p-k)\right)
\nonumber\\
                         &=& \sum_p \sum_{\pi} \left(\bar\psi^{\pi}(p), \sigma_3 \psi^{\pi}(p-k)\right)\ ,
\label{eq:4.2b}
\eea
\ese
and $\chis^{\text{L}}(h) = \chis^{\text L}(k=0,h)$.
Any nonanalytic dependence of $h$ will translate into corresponding nonanalytic dependences on the frequency,
wave number, or temperature via the scaling relations explained in Sec.~\ref{subsec:III.A}.\cite{zero_prefactor_footnote}

\subsection{The structure of diagrammatic perturbation theory}
\label{subsec:IV.A}

\begin{figure}[t]
\includegraphics[width=2.25cm]{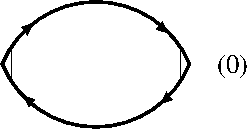}
\caption{The spin susceptibility to zeroth order in the interaction. The directed solid lines represent
               Green functions, the thin vertical lines represent the spin-density vertex $\sigma_3$.}
\label{fig:4}
\end{figure}

\begin{figure}[t]
\includegraphics[width=5.5cm]{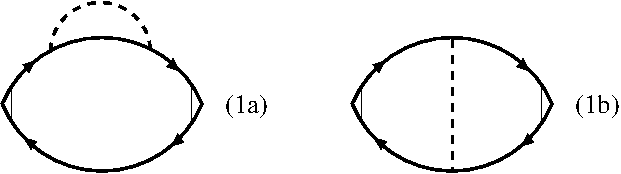}
\caption{The spin susceptibility to first order in the interaction. The dashed lines represent
              interaction amplitudes.}
\label{fig:5}
\end{figure}

\begin{figure}[t]
\includegraphics[width=8.5cm]{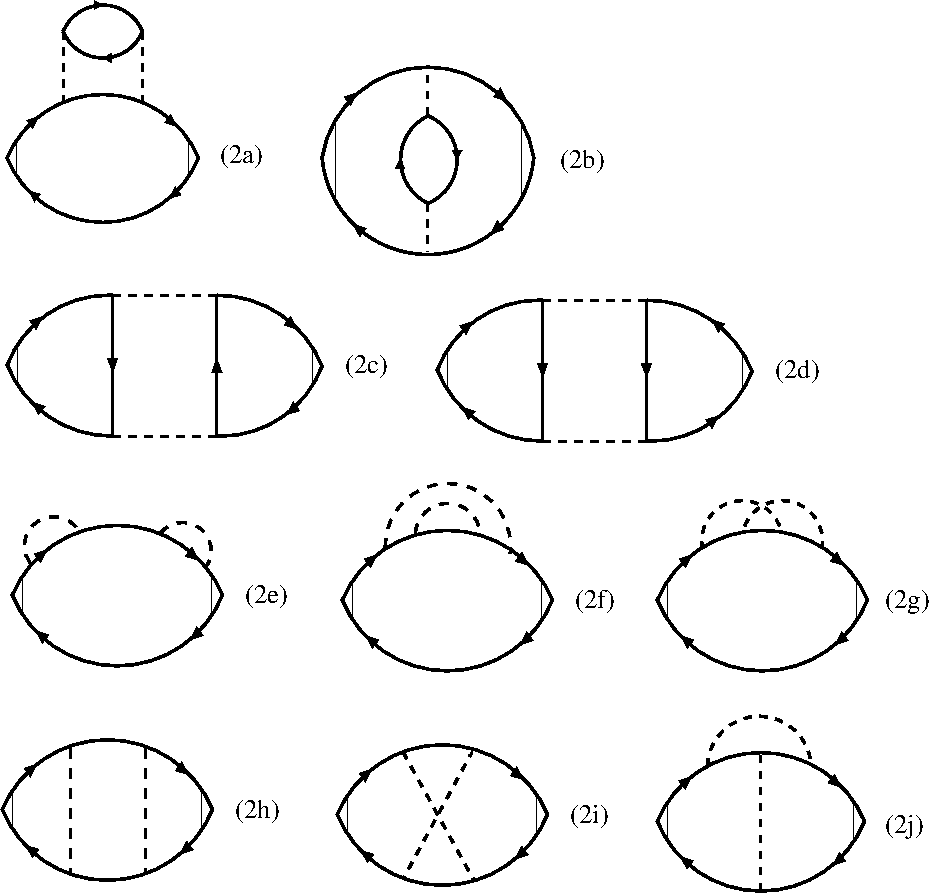}
\caption{The spin susceptibility to second order in the interaction. Diagrams (2a-d) represent
              1-loop integrals, diagrams (2e-j) represent two-loop integrals.}
\label{fig:6}
\end{figure}

Consider an expansion of $\chis$ in powers of the interaction amplitudes $\Gamma$. 
Standard diagrammatic perturbation theory leads us to consider the diagrams in Figs.~\ref{fig:4} - \ref{fig:6}. 
It is illustrative to perform a general structural analysis of the diagrams in each of these figures.

To zeroth order in the interaction $\chis$ is given by the simple fermion loop represented by diagram (0) in Fig.~\ref{fig:4}.
There is no frequency mixing, and this diagram has no hydrodynamic content. $\chis$ for
noninteracting electrons is therefore an analytic function of $h$ for any Fermi liquid, although even the noninteracting
system contains the relevant soft modes, see Sec.~\ref{subsec:III.A}. 

To first order in the interaction we have the two diagrams (1a,b) shown in Fig.~\ref{fig:5}. 
(A third diagram, which links two fermion loops by means of an interaction line with a zero momentum 
transfer vanishes due to charge conservation, and has no hydrodynamic content even if the external 
momentum is taken to be nonzero.) They do involve frequency mixing, but it is easy to see that they
cannot yield any nonanalyticities either. For a discussion of their structure, see Appendix~\ref{app:B}.

\begin{figure}[t]
\includegraphics[width=8.5cm]{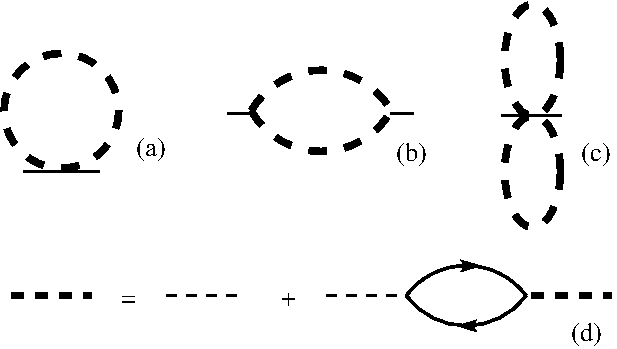}
\caption{(a) 1-loop integral in an effective field theory that contains the many-body diagrams
              (2a,b) in Fig.~\ref{fig:6}. (b) 1-loop integral that contains (2c,d) in Fig.~\ref{fig:6}.
              (c) 2-loop integral that contains (2e-j) in Fig.~\ref{fig:6}.
              (d) Thick dashed lines represent an RPA resummation of interaction amplitudes (thin
              dashed lines).}
\label{fig:7}
\end{figure}

To second order in the interaction, we have the diagrams shown in Fig.~\ref{fig:6}. (Again,
we do not show diagrams that contain interaction lines with a zero momentum transfer, which vanish
by charge conservation.) Structurally, these diagrams separate into two distinct groups. Diagrams
(2a) - (2d) correspond to integrals over one hydrodynamic wave vector, whereas diagrams (2e) - (2j)
correspond to two such integrations. While the two classes scale the same way, they
therefore are fundamentally different in a structural sense. This is reminiscent of the structure of a
loop expansion in an effective field theory. Indeed, for a Landau Fermi liquid there exists an effective
field theory that allows for a loop expansion, and the relation between the terms in the loop expansion
and (resummations of) many-body diagrams is known.\cite{Belitz_Kirkpatrick_2012a} An analogous
effective field theory can be constructed for a Dirac Fermi liquid,\cite{GDC_thesis} and the relation
between the diagrams in the field theory and in many-body theory, respectively, is summarized in Fig.~\ref{fig:7}. 
The most important aspect of the field theory and its loop-expansion structure is that it
allows for a renormalization-group (RG) analysis that guarantees that the functional form of any nonanalyticity 
obtained at a low-loop order cannot be changed by higher-loop contributions, only the prefactor can be
affected. The prescription for obtaining loop diagrams in the field theory from many-body diagrams is to
perform a random-phase-approximation (RPA) resummation of the interaction amplitudes (see Fig.~\ref{fig:7}(d)), 
consider the resulting screened amplitudes propagators, and contract all other electron Green functions to points.
Diagrams (2a,b) and (2c,d) then
correspond to the 1-loop diagrams shown in Fig.~\ref{fig:7}(a) and \ref{fig:7}(b), respectively, whereas diagrams (2e-j) 
correspond to the two-loop diagram shown in Fig.~\ref{fig:7}(c).
Note that diagrams (1a,b) also correspond to the one-loop diagram in Fig.~\ref{fig:7}(a), which
vanishes to linear order in the interaction. 

The conclusion from these considerations is that a systematic double expansion in the number
of integrations over the hydrodynamic momenta (equivalent to a loop expansion) and the number
of interaction amplitudes results in diagrams (2a - d) in Fig.~\ref{fig:6} as the lowest-order
contributions. There are terms of higher order in the loop expansion, some of which are also of
second order in the interaction (e.g., diagrams (2e - j) in
Fig.~\ref{fig:6}), but by the RG arguments mentioned above they cannot change the nature of 
any nonanalyticities. Similarly, at a given order in the loop expansion there are diagrams of
higher order in the interaction (e.g., any of the diagrams (2a - d) in Fig.~\ref{fig:6}) with the
dashed line replaced by the thick dashed line from Fig.~\ref{fig:7}(d)), but they scale the
same way as the lower-order ones and hence cannot change the nature of the 
nonanalyticity either. As a result, if diagrams (2a - d) in Fig.~\ref{fig:6} yield a nonanalytic contribution
to the spin susceptibility, then this result will be exact as far as the functional form of the
nonanalyticity is concerned. The prefactor of course will be perturbative. 

\subsection{Nonanalytic contributions to the spin susceptibility}
\label{subsec:IV.B}

We are now in a position to calculate the nonanalytic $h$-dependence of the spin susceptibility
$\chis$ to second order in the interaction amplitudes. We know from Ref.~\onlinecite{Kirkpatrick_Belitz_2019a}
that the leading nonanalyticity has the form $h^{d-1}$ in generic dimensions $d>1$, and $h^2\ln h$
in $d=3$, see Sec.~\ref{sec:I}, the only question is whether the prefactor is nonzero. (For comments
on the sign of the prefactor, see Sec.~\ref{subsubsec:IV.B.3} below.) Since the various interaction amplitudes are independent,
contributions from different interaction channels in Eq.~(2.14') cannot cancel each other. In order
to establish the existence of a nonzero prefactor, it therefore suffices to find one channel that
gives a nonzero result. 

Since the matrices $M^{\alpha\beta}$, Eqs.~(\ref{eqs:2.11}) are different in the limits $\Delta=0$ and
$\Delta \gg v\kF$, respectively, we need to distinguish between these two limits when writing contributions
to $\chis$ in terms of the quasiparticle resonances $F$. We start with $\Delta=0$, which was 
considered before in Ref.~\onlinecite{Kirkpatrick_Belitz_2019a}.

\subsubsection{The case $\Delta = 0$}
\label{subsubsec:IV.B.1}

\begin{figure}[t]
\includegraphics[width=8.5cm]{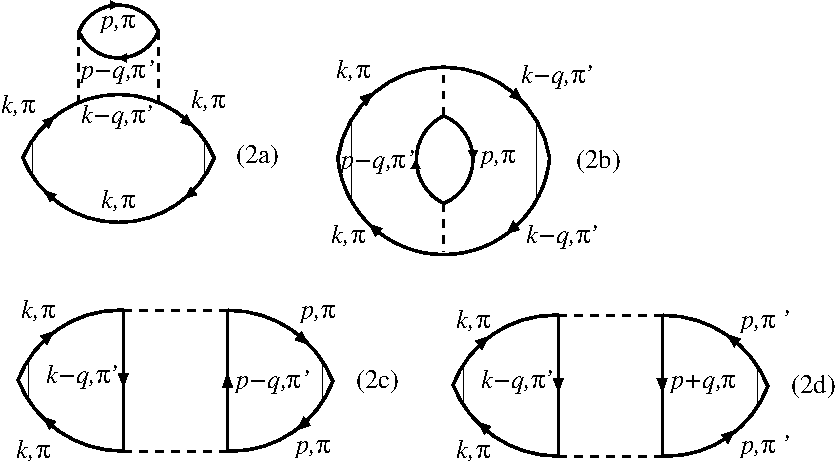}
\caption{1-loop contributions to the spin susceptibility to second order in the interaction amplitude $\Gamma_{\text{t},3}$.}
\label{fig:8}
\end{figure}

For the case $\Delta = 0$, it was shown in Ref.~\onlinecite{Kirkpatrick_Belitz_2019a}
that none of the spin-singlet amplitudes contribute, and neither do $\Gamma_{\text{t},1}$ or $\Gamma_{\text{t},2}$.
The remaining spin-triplet amplitudes, $\Gamma_{\text{t},3}$ and $\Gamma_{\text{t},4}$, do not mix to second order,
which leaves possible contributions of order  $\left(\Gamma_{\text{t},3}\right)^2$ and $\left(\Gamma_{\text{t},4}\right)^2$.
The contributions proportional to $\left(\Gamma_{\text{t},3}\right)^2$ 
were calculated in Ref.~\onlinecite{Kirkpatrick_Belitz_2019a}, and we quote the result in terms
of an integral:
\bea
\delta\chis^{\text L\,(3)}(h) &=& -2\left(\Gamma_{{\text t},3}\right)^2 \frac{\vF^2}{\vF^2 - v^2}  \sum_{q}{}^{'} \sum_{k,p}
   {\hat k}_z {\hat p}_z (1 - \hat{\bm k}\cdot\hat{\bm p})^2
\nonumber\\
&&\hskip 20pt \times \left(F_k^{++}\right)^2 F_{k-q}^{-+}\,\left(F_p^{+-}\right)^2 F_{p-q}^{--}\ .
\label{eq:4.3}
\eea
The corresponding diagrams are shown in Fig.~\ref{fig:8}; they are diagrams (2a - d) from Fig.~\ref{fig:6}
with explicit frequency-momentum and chirality labels. 
Note that diagrams (2a,b) lead to (4,2) partitions
of the six factors of $F$, but can be rewritten as (3,3) partitions in the case of a zero external momentum. 
All factors of $F$ in each of the two 3--$F$ convolutions have the same value of the cone index $\beta$, as is necessary for
the convolution to be a soft mode, see Sec.~\ref{subsec:III.B}. Importantly, however, that common value
of $\beta$ is different for the two 3--$F$ convolutions. That is,  $\Gamma_{{\text t},3}$ facilitates
inter-cone scattering only. Consequently, this contribution to $\chis$ will be
nonzero only if both cone indices contribute to the Fermi surface. For values of the coupling constant
$v$ that are on the order of the atomic velocity scale, this is not the case, see Fig.~\ref{fig:1}. 

\begin{figure}[t]
\includegraphics[width=8.5cm]{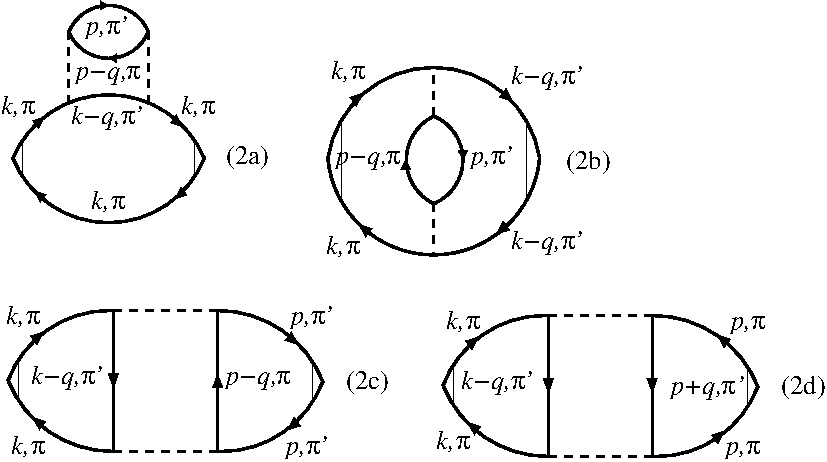}
\caption{Contributions to the spin susceptibility to second order in the interaction amplitude $\Gamma_{\text{t},4}$.}
\label{fig:9}
\end{figure}

If $\Delta=0$ due to a symmetry that enforces particle-number conversation for fermions with
a given chirality, then $\Gamma_{{\text t},4} = 0$ as well, see the remarks after Eq.~(2.14'). In that
case, $\chis$ has no nonanalytic contribution for sufficiently large $v$. This is the case considered
in Ref.~\onlinecite{Kirkpatrick_Belitz_2019a}. However, if $\Delta = 0$
due to fine tuning of the band structure, and not mandated by symmetry, then $\Gamma_{{\text t},4}$
will generically be nonzero and needs to be considered. The diagrams that lead to contributions
proportional to $(\Gamma_{\text{t},4})^2$ are shown in Fig.~\ref{fig:9}. 
A calculation along the same lines as the one leading to Eq.~(\ref{eq:4.3}) shows that in addition to an
inter-cone term, which has the same cone structure as Eq.~(\ref{eq:4.3}),  $\Gamma_{\text{t},4}$
facilitates an intra-cone scattering process where the two 3--$F$ convolution carry the same cone
index. For the latter we find
\bea
\delta\chis^{\text L\,(4)}(h) &=& 
 \left(\Gamma_{{\text t},4}\right)^2 \sum_{q}{}^{'} \sum_{k,p} {\hat k}_z {\hat p}_z (1 - \hat{\bm k}\cdot\hat{\bm p})^2  
\nonumber\\
&&\hskip -20pt \times  \sum_{\beta}  \left(F_k^{+\beta}\right)^2 F_{k-q}^{-\beta}\,\left(F_p^{+\beta}\right)^2 F_{p+q}^{-\beta}\ .
\label{eq:4.4}
\eea
This provides a nonanalytic contribution to $\chis$ even if only one cone index
contributes to the Fermi surface. 

The $F_k^{\alpha\beta}$ are given, for $\Delta=0$, by Eq.~(\ref{eq:2.9}) with ${\bm k}_{\Delta}$ from Eq.~(\ref{eq:2.4a}):
\be
F_k^{\alpha\beta} = \frac{1}{i\omega_n - \xi_{\bm k} - \beta \vert v{\bm k} - \alpha{\bm h}\vert}\ .
\label{eq:4.5}
\ee
Performing the integrals in Eqs.~(\ref{eq:4.3}) and (\ref{eq:4.4}) and extracting the leading nonanalytic
$h$-dependence we find
\bea
\delta\chis^{\text L}(h) &=& \frac{4}{5} \left[2(n_c - 1)\left(\NF \Gamma_{\text{t},3}\right)^2 + n_c \left(\NF\Gamma_{\text{t},4}\right)^2\right] 
\nonumber\\
 && \hskip 100pt \times  \frac{1}{V} \sum_{\bm q} \frac{1}{(\vF\vert{\bm q}\vert)^3}
\nonumber\\
&=& \NF\,\frac{1}{5}  \left[2(n_c - 1)\left(\NF \Gamma_{\text{t},3}\right)^2 + n_c \left(\NF\Gamma_{\text{t},4}\right)^2\right] 
\nonumber\\
&& \hskip 80pt \times (h/\epsilonF)^2 \ln(\epsilonF/h)\ .
\label{eq:4.6}
\eea
where we have specialized to $d=3$ in the second line. Here $n_c = 1,2$ is the number of cones that contribute to
the Fermi surface, and the term $\propto \left(\Gamma_{\text{t},4}\right)^2$ is present only if $\Gamma_{\text{t},4}$
is not zero by symmetry. This generalizes the result obtained in Ref.~\onlinecite{Kirkpatrick_Belitz_2019a}.

\subsubsection{The case $\Delta \gg v\kF$}
\label{subsubsec:IV.B.2}

In the limit $\Delta \gg v\kF$ we need to use the Green function with the matrices $M^{\alpha\beta}$ given by 
Eq.~(\ref{eq:2.11b}) instead of (\ref{eq:2.11a}). In this case, many interaction amplitudes contribute to the
nonanalyticity, in contrast to the $\Delta = 0$ case. In the light of the comments at the beginning of
Sec.~\ref{subsec:IV.B} we focus on the contributions from $\Gamma_{\text{t},3}$, as any nonzero
contribution suffices for our purposes. We furthermore consider only intra-cone scattering for the
up-cone ($\beta=+1$), which contributes to the Fermi surface irrespective of the value of $v$.
From Fig.~\ref{fig:8} we find
\bea
\delta\chis^{\text L\,(3,+)}(h) &=& 16 \left(\Gamma_{\text{t},3}\right)^2  \sum_{q}{}^{'}
\nonumber\\
&&\hskip -30pt \times \left[ \sum_{k} \left(F_k^{++}\right)^2 F_{k-q}^{-+} \sum_p \left(F_p^{++}\right)^2 F_{p-q}^{-+} \right.
\nonumber\\
&&\hskip -30pt \left. + 2 \sum_k \left(F_k^{++}\right)^3 F_{k-q}^{-+} \sum_p F_p^{++} F_{p-q}^{-+} \right]
\label{eq:4.7}
\eea

The functions $F$ in this case ($h, v\kF \ll \Delta$) are given by Eq.~(\ref{eq:2.9}) with ${\bm k}_{\Delta}$ from Eq.~(\ref{eq:2.4b}):
\be
F_k^{\alpha +} = \frac{1}{i\omega_n - \xi_{\bm k} - \Delta - \alpha h}\ .
\label{eq:4.8}
\ee
We see that, in this limit, the quasiparticle resonance $F_k^{\alpha +}$ is equal to the Green function for a Landau 
Fermi liquid, Eq.~(\ref{eq:2.12b}), with the Fermi energy shifted by $\Delta$ and $\alpha$ playing the role of the
spin projection index. The problem thus maps onto the corresponding one for a Landau Fermi liquid and we have,
for $d=3$,
\be
\delta\chis^{\text L\,(3,+)}(h) = 4\NF \left(\NF\Gamma_{\text{t},3}\right)^2 (h/\epsilonF)^2 \ln (\epsilonF/h)\ .
\label{eq:4.9}
\ee
We emphasize again that this is only a particular contribution to the nonanalytic behavior of $\chis$, there
are many others.

\subsubsection{Summary of nonanalytic contributions to $\chis$}
\label{subsubsec:IV.B.3}

Before we discuss the consequences of the nonanalytic behavior of $h$-dependence of $\chis^{\text{L}}$ for
magnetic quantum phase transitions, we summarize what we have concluded about this effect in 
various parameter regimes.

Basic scaling arguments imply that the soft modes discussed in Sec.~\ref{sec:III} lead to a
nonanalytic $h$-dependence of the form
\bea
\chis^{\text{L}} &=& \chis^{\text{L}}(h=0) 
\nonumber\\
&& + \chis^{(2)} \times \begin{cases} (h/\epsilonF)^{d-1}                      & \text{for $1<d<3$} \\
                                                                                   (h/\epsilonF)^2 \ln (\epsilonF/h) & \text{for $d=3$}
                                    \end{cases}
\nonumber\\
                            &&+ \text{analytic terms}\ .  
\label{eq:4.10}
\eea

The generic case is given by the full Hamiltonian in Eq.~(\ref{eq:2.1}) plus all of the interaction terms
shown in Eq.~(\ref{eq:2.14}) or (2.14'). In this case $\chis^{(2)}$ is always nonzero. There are several
contributions to $\chis^{(2)}$; in Sec.~\ref{subsubsec:IV.B.2} we have calculated a particular one, 
proportional to $(\Gamma_{\text{t},3})^2$, that is present irrespective of whether one or two cones 
contribute to the Fermi surface. The explicit calculation has been performed in the limit of a large gap, 
$\Delta \gg v\kF$. However, it is easy to see that an analogous term exists for $0 < \Delta < v\kF$, 
albeit with a small prefactor proportional to $(\Delta/v\kF)^2$. 

The case of a gapless Dirac metal $\Delta=0$, is special. There are fewer ways to generate a
nonanalyticity, and only the interaction constants  $\Gamma_{\text{t},3}$ and  $\Gamma_{\text{t},4}$
can contribute. These contributions have been calculated in Sec.~\ref{subsubsec:IV.B.1}.
If $\Delta=0$ accidentally, e.g., due to fine tuning of the system parameters, then 
$\Gamma_{\text{t},4}$ is generically nonzero and still assures that a nonanalyticity is present.
However, if $\Delta=0$ due to a symmetry that ensures particle-number conservation for each
chirality separately, then that same symmetry also implies that $\Gamma_{\text{t},4}=0$.
In this case $\Gamma_{\text{t},3}$ still leads to a nonanalyticity provided both cones
contribute to the Fermi surface. If the spin-orbit coupling parameter $v$ is a sizable
fraction of the atomic-scale velocity $v_0$, then this is not the case. In this case $\chis$
is an analytic function of $h$. Note that either $\Delta>0$ or $\Gamma_{\text{t},4}>0$
suffices for resurrecting the nonanalyticity. This is plausible, given that both of these
coupling constants break the same gauge symmetry in chirality space, see Sec.~\ref{sec:II}.

While our explicit calculations are perturbative with respect to the
electron-electron interaction, the scaling and RG arguments discussed in Sec.~\ref{subsec:IV.A}
imply that the results are indeed more general in the sense that the functional form of the
nonanalyticity is exact; only the prefactor $\chis^{(2)}$ is perturbative. Therefore, the only case that is
not necessarily robust against higher orders in a loop expansion (as defined in Sec.~\ref{subsec:IV.A})
is the null result for the case $\Delta = \Gamma_{\text{t},4}=0$ with only one cone contributing
to the Fermi surface. Here, we cannot exclude the possibility that higher orders in a loop expansion 
restore the missing coupling and lead to a nonzero prefactor of the nonanalyticity.

We finally comment on the sign of the prefactor of the nonanalyticity (see Ref.~\onlinecite{Brando_et_al_2016a},
where the same reasoning was given for a Landau Fermi liquid).  All of our explicit
calculations have yielded contributions for which $\chis^{(2)}>0$. This is not accidental.
The soft modes represent fluctuations that decrease the tendency of the Fermi liquid to order
magnetically, and hence decreases $\chis(h=0)$ with respect to its value in
the absence of the fluctuations. A magnetic field weakens these fluctuations (this can be seen
explicitly in Sec.~\ref{sec:III}, where $h$ gives the relevant soft modes a mass), and hence
leads to a positive correction to $\chis(h=0)$. The sign of $\chis^{(2)}$ is therefore
expected to be universal and positive.

\section{Magnetic quantum phase transition}
\label{sec:V}

We now discuss the consequences of the nonanalytic dependence of the spin susceptibility
on a magnetic field, Eq.~(\ref{eq:4.10}), for various magnetic quantum phase transitions. 

\subsection{Ferromagnets, canted ferromagnets, and ferrimagnets}
\label{subsec:V.A}

Consider the action $S_{\text{DM},h=0}$ for a Dirac metal, Eq.~(\ref{eq:2.15}), in zero field. Now assume
that the conduction electrons are subject to a fluctuating magnetization ${\bm m}(x)$.
The magnetization will then couple to the conduction electrons via a contribution $S_{\text{Z}}$
to the action that is of Zeeman form,
\be
S_{\text{Z}} = c \int dx\ {\bm m}(x)\cdot{\bm n}_{\text{s}}(x)\ ,
\label{eq:5.1}
\ee
where 
\be
{\bm n}_{\text{s}}(x) = \left(\bar\psi(x),({\bm\sigma}\otimes\pi_0)\psi(x)\right)
\label{eq:5.2}
\ee
is the vector generalization of Eq.~(\ref{eq:4.1b}), and $c$ is a coupling constant. This is
true irrespective of the origin of the magnetization; it may result from 
localized magnetic moments, or it may be itinerant in the sense that it is generated by
the conduction electrons.\cite{itinerant_footnote} Also, the ground state does not have to be a homogeneous
ferromagnet, e.g., a canted ferromagnet qualifies (for ferrimagnets, see below). For
what follows we only assume that the magnetization has a homogeneous component.
The action $S_{\text{DM}}[\bar\psi,\psi] + S_{\text{Z}}[{\bm m};\bar\psi,\psi]$ has to be augmented by 
a purely bosonic action ${\cal A}_{\text{OP}}[{\bm m}]$ that governs the behavior of the order-parameter field ${\bm m}$.
The partition function for the coupled fermion-boson system then reads (here we use the same notation and sign
convention as in Ref.~\onlinecite{Brando_et_al_2016a})
\bse
\label{eqs:5.3}
\bea
Z &=& \int D[{\bm m}]\,D[\bar\psi,\psi]\ e^{-{\cal A}[{\bm m}] + S_{\text{DM},h=0}[\bar\psi,\psi] + S_{\text{Z}}[{\bm m};\bar\psi,\psi]} 
\nonumber\\
\label{eq:5.3a}\\
&=& \int D[{\bm m}]\,e^{-{\cal A}_{\text{eff}}[{\bm m}]}\ ,
\label{eq:5.3b}
\eea
where
\bea
{\cal A}_{\text{eff}}[{\bm m}] &=& {\cal A}_{\text{OP}}[{\bm m}] 
\nonumber\\
&& - \ln \int D[\bar\psi,\psi]\,e^{S_{\text{DM},h=0}[\bar\psi,\psi] + S_{\text{Z}}[{\bm m};\bar\psi,\psi]}
\nonumber\\
\label{eq:5.3c}\\
&=& {\cal A}_{\text{OP}}[{\bm m}] - \ln \left\langle e^{S_{\text{Z}}[{\bm m};\bar\psi,\psi]} \right\rangle_{S_{\text{DM},h=0}}
\label{eq:5.3d}
\eea
\ese
is an effective bosonic action.  In Eq.~(\ref{eq:5.3d}) we have dropped a constant contribution to the effective action. 

The order-parameter action ${\cal A}_{\text{OP}}$ describes the magnetization in the absence of the coupling to the conduction
electrons. An appropriate choice for ${\cal A}_{\text{OP}}$ is thus a Landau-Ginzburg-Wilson action for the fluctuating magnetization.
The corresponding quantum phase transition was first studied by Hertz,\cite{Hertz_1976} who showed that the critical behavior is
mean-field-like. This is because the dynamical critical exponent $z=3$ lowers the upper critical dimension from 4 in the classical
theory to $d_c^+ = 4 - z = 1$ in the quantum case. It thus is plausible that it is a good approximation to replace the fluctuating
magnetization ${\bm m}(x)$ by its average $(0,0,m)$ even in the presence of the coupling to the conduction electrons.\cite{liquid_crystal_footnote}
The free-energy density $f$ then has a simple Landau form with a correction $\delta f$ due to the coupling to the fermionic
soft modes,
\bse
\label{eqs:5.4}
\be
f = t\,m^2 + u\,m^4 + \delta f(m)\ ,
\label{eq:5.4a}
\ee
with
\be
\delta f(m) = -(T/V) \ln \left\langle e^{\,cm\int dx\,n_{\text{s}}^3 (x)} \right\rangle_{S_{\text{DM}}}
\label{eq:5.4b}
\ee
\ese
Differentiating twice with respect to $m$ we obtain
\bea
\frac{d^2}{d m^2}\,\delta f(m) &=& \frac{-T}{V}\,c^2 \int dx\,dy\ \left\langle(\delta n_{\text{s}}^3(x)\,\delta n_{\text{s}}^3(y)\right\rangle_{S_{\text{DM},h=cm}}
\nonumber\\
&=& -c^2 \chis^{\text{L}}(h=cm)\ ,
\label{eq:5.5}
\eea
with $\chis^{\text{L}}(h=cm)$, as defined in Eq.~(\ref{eq:4.1a}), the longitudinal spin susceptibility of a Dirac metal in an effective
magnetic field $h=cm$. The contribution of the soft modes to the mean-field free-energy density is thus given by
\be
\delta f(m) = -c^2 \int_0^m dm' \int_0^{m'} dm''\ \chis^{\text{L}}(h=c m'')\ .
\label{eq:5.6}
\ee
With our result for $\chis^{\text{L}}(h)$, Eq.~(\ref{eq:4.10}), we finally obtain
\be
\delta f(m) = -{\tilde u} \times \begin{cases} m^{d+1} & \text{for $1<d<3$} \\
                                                                    m^4 \ln (1/m) & \text{for $d=3$}
                                              \end{cases}
\label{eq:5.7}
\ee   
for the nonanalytic contribution to $\delta f(m)$ (the analytic contributions merely redefine the Landau parameters in Eq.~(\ref{eq:5.4a})).                                           
The parameter $\tilde u$ is proportional to $\chis^{(2)}$ in Eq.~(\ref{eq:4.10}) and hence positive, see the discussion at the end of
Sec.~\ref{subsubsec:IV.B.3}. We see that $\delta f(m)$ provides, for all spatial dimensions $1<d<3$, a negative term in the mean-field 
free energy that dominates the $m^4$ term. This leads to a quantum phase transition that is necessarily first order.
The above derivation is the same as the one given for the ferromagnetic quantum phase transition in a Landau Fermi liquid in
Ref.~\onlinecite{Brando_et_al_2016a} and has been included here for completeness. In either case, the conclusion follows
from the result for the spin susceptibility in a magnetic field. 

In addition to homogeneous ferromagnetic order, the derivation obviously still holds for canted ferromagnets, i.e., a bipartite
lattice with ferromagnetic orders on each sublattice that are not collinear. It also holds for ferrimagnetic order, i.e., systems
with a fluctuating magnetization
\be
{\bm M}(x) = {\bm m}(x) + {\bm n}(x) \sum_{j=1}^N \cos({\bm k}_j\cdot{\bm x})\ .
\label{eq:5.8}
\ee
Here ${\bm m}(x)$ and ${\bm n}(x)$ are slowly fluctuating fields whose averages are the homogeneous magnetization
and the staggered magnetization, respectively, and the ${\bm k}_j$ are $N$ wave vectors that characterize the staggered
order. A ferrimagnet results if ${\bm m}$ and ${\bm n}$ acquire nonzero expectation values $m$ and $n$ at the same point in parameter
space. The dominant coupling between the conduction electrons and the magnetization will be to ${\bm m}$, since the
fermionic soft modes are soft at zero wave vector and frequency. The above derivation then still holds, and $n$ is simply
slaved to $m$. For Landau Fermi liquids this was discussed in detail in Ref.~\onlinecite{Kirkpatrick_Belitz_2012b}, and
the verbatim same reasoning holds in the case of a Dirac Fermi liquid. 

The only possible exception from the conclusion that the transition is first order is if $\Delta = \Gamma_{\text{t},4} = 0$.
In this case the spin susceptibility has no nonanalyticity to one-loop order, and by Eq.~(\ref{eq:5.6}) neither does $\delta f$.
If this result remains valid at higher-loop order, then the transition will be second order.

\subsection{Magnetic Nematics}
\label{subsec:V.B}

Magnetic nematics relate to ferromagnets the way non-s-wave superconductors relate to s-wave ones. 
Alternatively, the magnetic nematic transitions can be considered the spin-channel analogs of the Pomeranchuk instability in the
charge channel. They are characterized by an order parameter represented by a nonvanishing expectation value
\be
\left\langle\left({\bar\psi}(x),(\pi_0\otimes{\bm\sigma})f({\hat{\bm\nabla}}_{\bm x})\psi(x)\right)\right\rangle\ .
\label{eq:5.9}
\ee
Here $f$ is a tensor-valued monomial function of a vector variable, and $\hat{\bm\nabla}_{\bm x}$ denotes
the spatial gradient operator ${\bm\nabla}_{\bm x} = (\partial_x,\partial_y,\partial_z)$
divided by its norm (i.e., the Fourier transform of $\hat{\bm\nabla}_{\bm x}$ is $\hat{\bm k}$). 
The simplest case is a $p\,$-wave nematic, in which case $f$ is a vector-valued function and the order-parameter field
\be
N_{i}^{\alpha}(x) = \left({\bar\psi}(x),(\pi_0\otimes \sigma_i) \hat{\partial}^{\alpha} \psi(x)\right)
\label{eq:5.10}
\ee
carries a spin index $i$ and an orbital index $\alpha$. There are two distinct phases: An $\alpha$-phase where
$N_i^{\alpha} = N {\hat n}^{\alpha} {\hat N}_i$ with $\hat{\bm n}$  and
${\hat {\bm N}}$ unit vectors in orbital and spin space, respectively, and $N$ a scalar, and a $\beta$-phase, where 
$N_i^{\alpha} = N\delta_i^{\alpha}$.\cite{Wu_Zhang_2004} 

The quantum phase transition from a Landau Fermi liquid to a p\,-wave magnetic nematic phase has been studied
before. Reference~\onlinecite{Wu_et_al_2007} considered a theory analogous to Hertz's theory for the ferromagnetic
transition\cite{Hertz_1976} that treats the conduction electrons in a zero-loop approximation and yields a second-order
transition with mean-field critical behavior. In Ref.~\onlinecite{Kirkpatrick_Belitz_2011} it was shown that the 
electronic soft modes, which are neglected in Hertz theory, drive the transition first order in analogy to what happens
to the ferromagnetic transition. In what follows we show that the same conclusion holds for the quantum phase
transition from a Dirac Fermi liquid to a $p\,$-wave magnetic nematic phase. 

To this end, we recall the logic of the development in Secs.~\ref{sec:II}, \ref{sec:III}, \ref{sec:IV}, and \ref{subsec:V.A}.
(1) The eigenvalues $\lambda_{\bm k}$ of the single-particle Hamiltonian determine the quasiparticle resonances
$F_k$. (2) The convolutions of the $F_k$ describe two-particle excitations. The subset of massless
excitations, if any, constitutes the relevant soft modes. (3) If any of the soft modes are made massive by
the field conjugate to the order parameter (i.e., if they are soft of the first kind with respect to that field
in the nomenclature of Ref.~\onlinecite{Kirkpatrick_Belitz_2019a}), then the free energy must be a nonanalytic 
function of the conjugate field. (4) This implies, by virtue of the bilinear coupling between the conjugate field and the order
parameter, that the renormalized Landau free energy for the quantum phase transition is a nonanalytic 
function of the order parameter. This leads to a first-order transition, at least at the level of a renormalized
mean-field theory.

Before we apply this logic to the magnetic-nematic transition in a Dirac Fermi liquid, it is illustrative to reconsider
the Landau case.

\subsubsection{Landau case}
\label{subsubsec:V.B.1}

Let ${\mathfrak h}$ be a homogeneous field conjugate to the order-parameter field $N(x)$ (see Ref.~\onlinecite{Kirkpatrick_Belitz_2011}
for a discussion of how such a field can be realized by means of a non-homogeneous magnetic field). In general,
the Zeeman-like term in the single-particle Hamiltonian then reads $-{\frak h}^i_{\alpha}(\pi_0\otimes\sigma_i){\hat k}^{\alpha}$.
For simplicity, we consider the $\beta$-phase, where the Fermi-surface distortion is isotropic and the single-particle
Hamiltonian reads
\be
H_0 = \xi_{\bm k}(\pi_0\otimes\sigma_0) - {\frak h}(\pi_0\otimes{\bm\sigma})\cdot{\hat{\bm k}}\ ,
\label{eq:5.11}
\ee
with ${\frak h}$ a scalar field. (For the $\alpha$-phase analogous arguments apply, but the anisotropy 
makes the development more cumbersome.) The eigenvalues are
\bse
\label{eqs:5.12}
\be
\lambda_{\bm k}^{\alpha} = \xi_{\bm k} - \alpha{\frak h}\ .
\label{eq:5.12a}
\ee
and the single-particle resonances are
\be
F_k^{\alpha} = \frac{1}{i\omega_n - \lambda_{\bm k}^{\alpha}} = \frac{1}{i\omega_n - \xi_{\bm k} + \alpha{\frak h}}\ .
\label{eq:5.12b}
\ee
\ese
Here $\alpha=\pm$ labels the two sheets of the Fermi surface that is split by ${\frak h}$, and each eigenvalue is 
two-fold degenerate due to the $\pi_0$ matrix that is redundant in the Landau case. Note that the field ${\frak h}$
introduces a spin-orbit coupling, and a chirality degree of freedom, even in the Landau case. That is, a
spin-orbit coupling is spontaneously generated by the nematic magnetic order, irrespective of whether the
order is induced by ${\frak h}$ or spontaneous.\cite{Wu_Zhang_2004}

We see that the field ${\frak h}$ splits the Fermi surface in formally the same
way as a physical magnetic field. That is, the eigenvalues of the Hamiltonian, and hence the single-particle
resonances, are the same as for the case of a physical magnetic field $h$ coupling to the Landau
Fermi liquid, see Eq.~(\ref{eq:2.12b}). The further development is now obvious: 
The soft modes are given by Eqs.~(\ref{eqs:3.2}) with $h$ replaced by
${\frak h}$ and $\sigma_{1,2}$ by $\alpha_{1,2}$, and the modes with $\alpha_1 \neq \alpha_2$ are soft
of the first kind with respect to ${\frak h}$. The renormalized mean-field free energy therefore has a
nonanalytic contribution that is given by Eq.~(\ref{eq:5.7}) with $m$ replaced by $N$, and the 
quantum phase transition is first order. For the $\alpha$-phase one obtains the same structure, but
the field term in the denominator of the soft mode, Eq.~(\ref{eq:3.2b}), has an angular dependence
related to the angle between the two unit vectors $\hat{\bm N}$ and $\hat{\bm n}$ that characterize
the order parameter. 

This is the result that was first obtained in Ref.~\onlinecite{Kirkpatrick_Belitz_2011}
by means of arguments that were technically more involved. 

\subsubsection{Dirac case}
\label{subsubsec:V.B.2}

Now consider the Dirac case. The Hamiltonian is
\bea
H_0 &=& \xi_{\bm k} (\pi_0\otimes\sigma_0) + v(\pi_3\otimes{\bm\sigma})\cdot{\bm k} +\Delta(\pi_1\otimes\sigma_0) 
\nonumber\\
&&  - {\frak h}(\pi_0\otimes {\bm\sigma})\cdot{\hat{\bm k}} \ .
\label{eq:5.13}
\eea
Solving the eigenvalue problem yields
\be
\lambda_{\bm k}^{\alpha\beta} = \xi_{\bm k} + \beta \vert v{\bm k}_{\Delta} - \alpha{\bm{\mathfrak h}}\vert\ ,
\label{eq:5.14}
\ee
with $\bm{\mathfrak h} = (0,0,{\mathfrak h})$, ${\bm k}_{\Delta}$ from Eq.~(\ref{eq:2.3}), and $\alpha,\beta = \pm 1$. 
We see that, as in the Landau case,
the eigenvalues of the Hamiltonian (\ref{eq:5.13}) map onto those of the corresponding Hamiltonian in a
physical magnetic field ${\bm h}$, Eqs.~(\ref{eq:2.1}, \ref{eq:2.2a}). As a result, the soft-mode structures are the same,
all of the results from Secs.~\ref{sec:IV} and \ref{subsec:V.A} still apply, and in particular the quantum phase
transition from a Dirac Fermi liquid to a $p\,$-wave (Dirac) magnetic nematic is first order. Again, the only
possible exception is the case where both $\Delta$ and $\Gamma_{\text{t},4}$ vanish due to a symmetry,
see the remarks at the end of Sec.~\ref{subsec:V.A}.

\section{Discussion and Conclusion}
\label{sec:VI}

Here we elaborate on various discussions given so far, discuss points that we have not covered yet, and conclude 
with a summary of our results.

\subsection{Discussion}
\label{subsec:VI.A}

\subsubsection{Soft two-particle excitations}
\label{subsubsec:VI.A.1}

An important concept underlying our discussion are the soft two-particle, or four-fermion, excitations that were derived in 
Sec.~\ref{sec:III}. They must be distinguished from the single-particle excitations, described by the Green function, which
in clean fermion systems are also soft. Their very different natures can be seen, for instance, from the fact that the 
single-particle excitations become massive in the presence of quenched disorder, whereas the two-particle excitations
remain soft, albeit with a diffusive energy-momentum relation rather than a ballistic one. Of crucial importance for the effect 
of the two-particle soft modes on any quantum phase transition is how they are affected by the external field
conjugate to the order parameter. There are two possibilities: (1) The soft modes are made massive by the conjugate field,
i.e., they are soft modes of the first kind with respect to the field in the nomenclature of Ref.~\onlinecite{Kirkpatrick_Belitz_2019a}.
Then the free energy and all of its derivatives with respect to the field must be nonanalytic functions of the field. Since, 
in the ordered phase, the nonvanishing order parameter is seen as an effective field by the conduction electrons, the 
mean-field free-energy functional must then be a nonanalytic function of the order parameter. This leads to a quantum 
phase transition that is generically first order. The quantum ferromagnetic transition in an ordinary metal is an example 
of this case: The soft modes in the transverse spin-triplet channel given by Eq.~(\ref{eq:3.2b}) with $\sigma_1 \neq \sigma_2$
are made massive by a magnetic field. $h$. Consequently, the spin susceptibility is a nonanalytic function of $h$ (and by
scaling, also of the wave number $k$ and the temperature $T$; for a review of the long history of this topic see
Ref.~\onlinecite{Brando_et_al_2016a}), and
the ferromagnetic quantum phase transition is first order. In the current paper we have shown that the same conclusion
holds in a Dirac metal, and also for the nematic magnetic transition in both Dirac and Landau metals.
(2) The soft modes remain soft in the presence of the conjugate field, i.e., they are soft of the second kind with respect
to the field. In this case the relevant susceptibility is an analytic function of the field and the quantum phase transition
is generically second order. As an example, the density susceptibility $\partial n/\partial\mu$ is an analytic function of
$\mu$, $k$, and $T$, and the electronic nematic quantum phase transition in the charge channel is second order.

This concept of two classes of soft modes, and its consequences for quantum phase transitions, was first discussed
in Ref.~\onlinecite{Belitz_Kirkpatrick_Vojta_2002}.

A very helpful aspect of the two-particle soft modes in a Fermi liquid, Landau or Dirac, is that their nature can be completely 
determined by considering the respective Fermi gas, as was demonstrated in Sec.~\ref{sec:III}. The reason why
the electron-electron interaction cannot change their structure (unless it is strong enough to destroy the Fermi liquid)
can be understood in various ways. One is to invoke Fermi-liquid theory. Another is the realization that the frequency
structure of any electron-electron interaction consists, by time translational invariance, of one delta-function constraint
for four fermionic frequencies (in Eq.~(\ref{eq:2.14}) this constraint has already been eliminated in favor of three
independent frequencies.) As a result, any interaction contribution to the soft-mode denominator in Eq.~(\ref{eq:3.2b})
or (\ref{eq:3.4c}) will necessarily carry a frequency $i\Omega$ and thus cannot give the soft mode a mass. A third
way relies on more formal arguments that interpret the soft modes as the Goldstone modes of a spontaneously
broken symmetry in Matsubara frequency space, i.e., the symmetry between retarded and advanced degrees of
freedom. This symmetry is broken whenever the single-particle spectrum is nonzero, irrespective
of the interaction. This interpretation of the soft modes was first given by Wegner for noninteracting disordered 
systems;\cite{Wegner_1979} it was later generalized to interacting systems with \cite{Belitz_Kirkpatrick_1997, Kirkpatrick_Belitz_2002}
or without \cite{Belitz_Kirkpatrick_2012a} quenched disorder. 

We also stress again that our considerations are independent of the topological properties of the respective
systems. Consider, for instance, the spectrum shown in the second panel of Fig.~\ref{fig:1}. Depending on
the parameter values, there may or may not be surface states that connect the two bands, and this is crucial for
 whether or not the system has nontrivial topological properties.\cite{Zhang_et_al_2009, Delta_footnote}
By contrast, the soft-mode spectrum is the same in either case, and so is the nature of the quantum magnetic
transition, if any.

\subsubsection{Validity of the theory for various magnetic transitions}
\label{subsubsec:VI.A.2}

We stress the generality of the derivation given in Sec.~\ref{subsec:V.A}, which follows the reasoning first
given in Ref.~\onlinecite{Brando_et_al_2016a}. The {\em only} requirement is that the magnetization has
a nonvanishing homogeneous component. As a result, the quantum phase transition is first order for
canted ferromagnets and for ferrimagnets as well as for homogeneous ferromagnets. For the Landau
case this was first realized in Ref.~\onlinecite{Kirkpatrick_Belitz_2012b}, and it holds true in the Dirac
case as well. It also holds irrespective of whether the magnetization is due to localized moments or
the conduction electrons themselves, see Ref.~\onlinecite{itinerant_footnote}.

For the magnetic nematic quantum phase transition a separate analysis is necessary, since the homogeneous 
magnetization vanishes and the conjugate field is not the physical magnetic field. In Sec.~\ref{subsec:V.B}
we showed that the soft-mode structure renders this transition first order as well. For the Landau
case this conclusion was first reached in Ref.~\onlinecite{Kirkpatrick_Belitz_2011}, but the current
derivation drastically simplifies the derivation.

\subsubsection{The role of spatial inversion symmetry}
\label{subsubsec:VI.A.3}

An important aspect of the Hamiltonian underlying a Dirac metal, Eq.~(\ref{eq:2.1}), is that it is
invariant under spatial inversion in addition to time reversal. From a pure symmetry point of view,
invariance under spatial inversion is what requires the existence of the chiral degree of freedom,
as $\bm\sigma\cdot\bm k$ is invariant under time reversal, but not under spatial inversion. Any system
with a strong spin-orbit coupling and a space group that lacks inversion symmetry will behave drastically 
differently from what we have described in the current paper. This physical situation will be discussed
elsewhere.\cite{us_unpublished}

\subsection{Summary, and Conclusion}
\label{subsec:VI.B}

In summary, we have considered the ferromagnetic quantum phase transition in a Dirac metal,
defined as an interacting electron system whose single-particle Hamiltonian is given by Eq.~(\ref{eq:2.1})
with a chemical potential $\mu > 0$. In such a system the spin-orbit interaction renders massive the
soft modes that drive the transition first order in an ordinary metal. However, we have shown that
the chirality degree of freedom leads to a new class of soft modes that also couple to the ferromagnetic
order parameter and again lead to a first-order quantum phase transition, contrary to what one might
naively expect. The same conclusion holds for canted ferromagnets, ferrimagnets, and magnetic
nematics. The chiral nature of the conduction electrons in a Dirac metal is crucial for this
conclusion. In systems with broken spatial inversion symmetry a strong spin-orbit interaction will
give the relevant soft modes a mass, but the absence of the chirality degree of freedom means
that no new soft modes are generated. In such systems we expect a ferromagnetic quantum
critical point if the spin-orbit interaction is strong enough.

\acknowledgments

This work was initiated at the Telluride Science Research Center (TSRC). 
We thank George de Coster and Hisashi Kotegawa for discussions.

\bigskip
\appendix

\section{Exact single-particle Green function}
\label{app:A}

The single-particle Green function $G$, Eq.~(\ref{eq:2.8}), can be expressed in terms of the quasiparticle resonances $F$, Eq.~(\ref{eq:2.9}),
as follows:

\be
G_k = \prod_{\alpha,\beta=\pm} F_k^{\alpha,\beta} \sum_{i,j=0,3} g_k^{ij}\left(\pi_i\otimes\sigma_j\right)
\label{eq:A.1}
\ee
with
\bse
\label{eqs:A.2}
\bea
g_k^{00} &=& (i\omega_n - \xi_{\bm k})\left[(i\omega_n - \xi_{\bm k})^2 - v^2{\bm k}^2 - h^2 - \Delta^2\right]\ ,
\label{eq:A.2a}\\
g_k^{01} &=& -2 h v^2 k_z k_x\ ,
\label{eq:A.2b}\\
g_k^{02} &=& - 2 h v^2 k_z k_y\ ,
\label{eq:A.2c}\\
g_k^{03} &=& -h \left[(i\omega_n - \xi_{\bm k})^2 + v^2 k_z^2 - v^2 (k_x^2 + k_y^2) - h^2 + \Delta^2\right]\ ,
\nonumber\\
\label{eq:A.2d}\\
g_k^{10} &=& \Delta \left[ (i\omega_n - \xi_{\bm k})^2 - v^2 {\bm k}^2 + h^2 - \Delta^2 \right]\ ,
\label{eq:A.2e}\\
g_k^{11} &=& g_k^{12} = 0 \ ,
\label{eq:A.2f}\\
g_k^{13} &=& -2\Delta (i\omega_n - \xi_{\bm k}) h\ ,
\label{eq:A.2g}\\
g_k^{20} &=& g_k^{23} = 0\ ,
\label{eq:A.2h}\\
g_k^{21} &=& 2\Delta h v k_y\ ,
\label{eq:A.2i}\\
g_k^{22} &=& -2\Delta h v k_x\ ,
\label{eq:A.2j}\\
g_k^{30} &=& -2 (i\omega_n - \xi_{\bm k}) h v k_z\ ,
\label{eq:A.2k}\\
g_k^{31} &=& \left[ (i\omega_n - \xi_{\bm k})^2 - v^2{\bm k}^2 - h^2 - \Delta^2\right] v k_x\ ,
\label{eq:A.2l}\\
g_k^{32} &=& \left[ (i\omega_n - \xi_{\bm k})^2 - v^2{\bm k}^2 - h^2 - \Delta^2\right] v k_y\ ,
\label{eq:A.2m}\\
g_k^{33} &=& \left[ (i\omega_n - \xi_{\bm k})^2 - v^2 {\bm k}^2 + h^2 - \Delta^2\right] v k_z\ .
\eea
\ese
Note that the product of the four $F_k$, which is the determinant of the inverse Green function, is independent of the
factor $s_{\bm k}$ in Eq.~(\ref{eq:2.3}).

\section{Absence of nonanalyticities to first order in the interaction}
\label{app:B}

\begin{figure}[b]
\includegraphics[width=5.5cm]{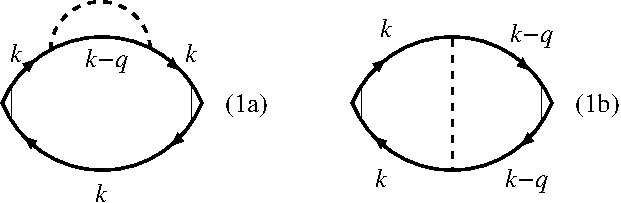}
\caption{The frequency-momentum structure of the first-order contributions to the spin susceptibility.}
\label{fig:B.1}
\end{figure}

Here we discuss the soft-mode content of the perturbative contributions to $\chis$ to first order in the interaction,
diagrams (1a,b) in Fig.~\ref{fig:5}. Consider diagram (1a), which is shown again in Fig.~\ref{fig:B.1} with the
frequency-momentum labels added. 
Within straightforward many-body perturbation theory, all bosonic frequencies included in the frequency-momentum
label $q$ are summed over. Ignoring for now the wave-number restriction inherent in the interaction amplitude, see
Eq.~(\ref{eq:2.14}) and the related remarks, the structure of diagram (1a) is
\bse
\label{eqs:B.1}
\bea
(1a) \propto \frac{T}{V} \sum_q \sum_k G_k^3\,G_{k-q} &=& \frac{T}{V}\sum_q G_q \sum_k G_k^3
\nonumber\\
&=& n \sum_k G_k^3
\label{eq:B.1.a}
\eea
with $n = (T/V)\sum_q G_q$ the particle-number density. The wave-number restriction just replaces the
Fermi wave number $\kF$ in $n$ by the UV cutoff $\Lambda$. The diagram thus amounts to a constant times an
integral that has no hydrodynamic content. The same is true for diagram (1b), which has the structure
\be
(1b) \propto \frac{T}{V} \sum_q \sum_k G_k^2\,G_{k-q}^2 \propto \frac{\partial n}{\partial\mu} \sum_k G_k^2\ .
\label{eq:B.1b}
\ee
\ese
Since neither $n$ nor $\partial n/\partial\mu$ can be nonanalytic functions of $h$, for reasons discussed
in Ref.~\onlinecite{Kirkpatrick_Belitz_2019a}, the same is true for these contributions to $\chi_s$.

Now consider a calculation that focuses entirely on soft modes and hydrodynamic content. Within
perturbation theory this can be done by restricting the frequency part of the sum over $q = (i\Omega,{\bm q})$
to frequencies for which $(\omega - \Omega)\omega < 0$ (here $i\omega$ is the frequency component of
$k$). In an effective field theory that focuses entirely on soft modes and ignores massive ones, or integrates 
them out in a simple approximation, this restriction is built into the structure of the theory; see Ref.~\onlinecite{Belitz_Kirkpatrick_2012a}
for an example. In such theories the analytic nature of the first-order contributions is
less obvious. However, by the above arguments they represent analytic contributions with a subtraction
that must be analytic, and hence they must be analytic themselves. Explicit calculations within such
frameworks confirm this.

We finally mention that in the presence of quenched disorder this structure gets modified, and there is
a nonanalytic contribution to $\chis$ at first order in the interaction.\cite{Altshuler_Aronov_1984}


\begin{thebibliography}{53}
\bibitem[{Del()}]{Delta_footnote}
\bibinfo{note}{In general, the gap parameter $\Delta$ is an even function of
  $\bm k$. This ${\bm k}$-dependence is important for the topological
  properties of the sytem,\cite{Zhang_et_al_2009} but is is not of qualitative
  importance for our purposes, and we neglect it.}
\bibitem[{cry()}]{crystal_field_footnote}
\bibinfo{note}{Crystal-field effects break this rotational invariance in spin
  space. They are generically much weaker than the spin-orbit effects we take
  into account, and we neglect them.}
\bibitem[{iti()}]{itinerant_footnote}
\bibinfo{note}{In the case of an itinerant magnet one might wonder whether the
  spin-triplet interaction has to be eliminated from $S_{\text{DM}}$, since it
  is necessary for producing the magnetization. This is not the case, since a
  spin-triplet interaction will always be generated under renormalization as
  long as a spin-singlet interaction is present. The action for the conduction
  electrons in the itinerant case is therefore still given by $S_{\text{DM}}$,
  albeit with a a renormalized spin-triplet interaction amplitude that is
  smaller than the threshold value necessary for producing a magnetization in
  the isolated fermion system.}
\end{thebibliography}

\end{document}